\documentclass[10pt,journal,compsoc]{IEEEtran}
\ifCLASSOPTIONcompsoc
  \usepackage[nocompress]{cite}
\else
  \usepackage{cite}
\fi

\ifCLASSINFOpdf
  \usepackage[pdftex]{graphicx}
  \DeclareGraphicsExtensions{.pdf,.jpeg,.png}
\else
  \usepackage[dvips]{graphicx}
  \DeclareGraphicsExtensions{.eps}
\fi

\usepackage{amsmath}
\usepackage{amsfonts,amssymb}
\usepackage{array}
\usepackage[11pt]{moresize}
\newtheorem{remark}{Remark}[section]

\usepackage{stfloats}

\usepackage{url}
\usepackage{acronym}
 
\usepackage{enumitem}
\usepackage[usenames, dvipsnames]{color}

\begin{document}
\acrodef{MGF}[MGF]{Moment Generating Function}
\acrodef{PDF}[PDF]{Probability Distribution Function}
\acrodef{CDF}[CDF]{Cumulative Distribution Function}
\acrodef{RV}[RV]{Random Variable}
\acrodef{PPP}[PPP]{Poisson Point Process}
\acrodef{BPP}[BPP]{Binomial Point Process}
\acrodef{RWPM}[RWPM]{Random Waypoint Mobility}
\acrodef{MAC}[MAC]{Medium Access Control}
\acrodef{SINR}[SINR]{Signal-to-Interference-and-Noise-Ratio}
\acrodef{i.i.d.}[i.i.d.]{independent and identically distributed}
%
\title{Temporal Correlation of Interference and Outage in Mobile Networks over One-Dimensional Finite Regions}
%
%
%

\author{Konstantinos Koufos and Carl P. Dettmann 
\IEEEcompsocitemizethanks{\IEEEcompsocthanksitem The authors are with the School of Mathematics,  University of Bristol, BS8 1TW, Bristol, United Kingdom. 
E-mail: \{K.Koufos, Carl.Dettmann\}@bristol.ac.uk }
}

\IEEEtitleabstractindextext{%
\begin{abstract}
In practice, wireless networks are deployed over finite domains, the level of mobility is different at different locations, and user mobility is correlated over time. All these features have an impact on the temporal properties of interference which is often neglected. In this paper, we show how to incorporate correlated user mobility into the interference and outage correlation models. We use the random waypoint mobility model over a bounded one-dimensional domain as an example model inducing correlation, and we calculate its displacement law at different locations. Based on that, we illustrate that the temporal correlations of interference and outage are  location-dependent, being lower close to the centre of the domain, where the level of mobility is higher than near the boundary. Close to the boundary, more time is also needed to see uncorrelated interference. Our findings suggest that an accurate description of the mobility pattern is important, because it leads to more accurate understanding/modeling of interference and receiver performance.
\end{abstract}

}

\maketitle

\IEEEdisplaynontitleabstractindextext

%
\IEEEpeerreviewmaketitle

\IEEEraisesectionheading{\section{Introduction}\label{sec:introduction}}

%
%
%
%
\IEEEPARstart{T}{he} performance of wireless networks is limited by interference. Interference is correlated over time when there are temporal correlations in the propagation channel, the user traffic and the user location~\cite{Ganti2009}. Interference correlation is directly related to the correlation of outage and because of that, it also affects other network performance metrics, e.g., the temporal diversity gain, the multi-hop delay, etc., thus becoming essential in the design of routing protocols, retransmission and \ac{MAC} schemes. The temporal correlation of interference has received some attention in the literature, however, under the assumption of infinite networks, where the locations of interferers are modeled by a \ac{PPP} in the infinite plane~\cite{Ganti2009,Haenggi2013a,Schilcher2011, Haenggi2010,Haenggi2013b, Schilcher2014,Gong2011,Gong2014,Gong2013}. Also, the impact of mobility on the interference correlation has been studied for mobility models which do not introduce correlation in the locations of a user over time~\cite{Haenggi2010,Haenggi2013b, Schilcher2014, Gong2011, Gong2014,Gong2013}. 

In~\cite{Ganti2009}, the interference correlation is investigated for static Poisson networks, with fixed but unknown  user locations, and slotted ALOHA.  In that case, the interferers in every time slot are selected from the same set of users, making the locations of interferers and subsequently interference pattern correlated over time. It is shown in~\cite{Ganti2009} that a high transmission probability is associated with a high interference correlation, and the temporal diversity gain due to retransmissions may completely vanish~\cite{Haenggi2013a}. The correlation becomes higher in a block fading channel and also, under correlated user traffic~\cite{Schilcher2011}. 

If the locations of interferers in each time slot are drawn from a new and independent realization of the \ac{PPP}, an infinite user velocity is essentially modeled~\cite{Haenggi2010}. With infinite velocity, the local delay, i.e., the mean time needed to connect to the nearest neighbor, is always finite~\cite{Haenggi2010,Haenggi2013b} while, in the static case, it may have a heavy tail. In relay chains under a Poisson field of interferers, the mean and variance of the end-to-end delay become smaller, when the locations of interferers are uncorrelated over time~\cite{Schilcher2014}. 

The impact of mobility on the interference correlation is studied in~\cite{Gong2011} for Poisson networks and various mobility models, i.e., constrained \ac{i.i.d.} mobility, random walk and discrete-time Brownian motion. For all the considered models, the uniform distribution of users is preserved between the time slots. It is shown in~\cite{Gong2011} that the interference correlation decreases inversely proportional with the mean speed of the users. In~\cite{Gong2014}, the results are extended to illustrate that the interference and outage are positively correlated, when the desired transmitter is placed at a fixed and known distance from the receiver.  Different levels of mobility are studied in~\cite{Gong2013}, where it is assumed that in each time slot, only a fraction of users remains static, and the rest move according to a mobility model preserving uniformity. A higher fraction of mobile users is associated with a lower correlation in the outage probability resulting in higher diversity gains~\cite{Gong2013}.

In practice, networks have finite boundaries and the concept of a typical receiver which is placed at the origin and where the network performance is assessed is not always realistic. This issue has already been pointed out in~\cite{Gong2014,Orestis2015,Pratt2015,Banani2015}, where single-snapshot analysis of interference and outage are carried out at different locations. For a Poisson network deployed over a convex domain, the receivers close to the boundaries experience more outage due to isolation~\cite{Coon2012}, but less due to interference~\cite{Orestis2015}. The location of the receiver  becomes more important in non-uniform deployments, where the interference would naturally vary more. In mobile networks following the \ac{RWPM} model, see for instance~\cite{Paolo2003, Bettstetter2004, Esa2004, Esa2006}, the users tend to concentrate close to the center of the area. Because of that, the interference over there becomes significantly higher than at the borders, motivating the use of location-aware routing protocols and \ac{MAC} schemes~\cite{Pratt2015}. Under the \ac{RWPM} model, the mean interference at the center becomes at least twice the mean interference generated by the equi-dense \ac{PPP}~\cite{Gong2014}. 

The temporal correlation of interference and outage for networks with finite boundaries and correlated mobility is yet to be studied. If the users are distributed according to a non-homogeneous \ac{PPP} and remain static, the analysis in~\cite{Ganti2009} still holds, i.e., the correlation coefficient will be  proportional to the random access probability, and inversely proportional to the second moment of the fading \ac{RV}. In addition, if the user locations are \ac{i.i.d.} over the time slots, the temporal correlation of interference is lower at the boundary than at the center~\cite{Koufos2017}. In this paper, we take a step further and consider the case where the mobility induces correlation in the user location over time. We will show how to incorporate the correlated mobility into the interference model, and we will illustrate that the temporal correlations of interference and outage are in general location-dependent. We will examine how quickly the interference correlation decays at different locations. Note that we do not treat group mobility models, where the locations of different users are correlated~\cite{Bai2004}. 

We need a mobility model which is defined over a finite area and it introduces temporal correlation in the user location. The \ac{RWPM} model has both of these features and it has been widely-used in the performance assessment of mobile wireless ad hoc networks, see for instance~\cite{Gong2014, Marina2001, Yang2012, Lassila2006}  as well as  cellular networks, e.g., in~\cite{Bettstetter2004, Esa2007}. The stationary node distribution and stochastic properties of the \ac{RWPM} has been studied in~\cite{Paolo2003, Bettstetter2004, Esa2004} over a line segment, rectangular and circular areas, then extended in~\cite{Esa2006} at an arbitrary convex domain, also with non-uniform waypoint distribution. In~\cite{Esa2005}, the stationary node distribution with \ac{RWPM} has also been derived for the unit hypersphere.

In the literature, the temporal correlation of interference in mobile networks has been studied in the  continuous one- and two-dimensional space. The correlation is described only between two time instances, and the mobility, as discussed above, is modeled either by a new and independent realization of the \ac{PPP} for some of the users or by a uniform displacement. Since the \ac{RWPM} model exhibits correlated and location-dependent mobility, and it is also defined in a bounded area, getting the user displacement law in continuous domain becomes challenging. In this paper, we have identified the \ac{PDF} of the user displacement in the one-dimensional finite lattice and up to three time instances for a positive think time. Note that the discretization of space is not a limitation of the model per se, because the lattice could be densified at an arbitrary order, provided that the displacement law is available, and continuous-domain approximation can be derived. In this paper, we have approximated the displacement law for a large number of time instances, thus obtain continuous-domain approximations, only for zero think time. 

Even though considering a discretized version of the \ac{RWPM} model in the one-dimensional space may seem to be overly simplistic, it suffices to illustrate how to incorporate the mobility correlations into the interference model, and allows us to get an insight into the interference and outage statistics which is not available with the existing models in the literature. In addition, we will show how to obtain continuous-domain approximations by densifying the lattice. Other mobility models over one-dimensional lattice can be treated in the same way, provided that the user distribution and displacement law are available. One-dimensional network models may also find practical applications, for instance, in vehicular networks. Obtaining the user displacement law over two-dimensional deployment areas has been left as future topic to study, but the main outcomes of this paper will still hold. 

To motivate a little bit further the use of \ac{RWPM} model, we point out that the model allows studying: (i) different levels of mobility by varying the think time, and (ii) a static network with a uniform user distribution in the limit of infinite think time. The contributions of this paper are:
\begin{itemize}[leftmargin=.25in]
\item For the \ac{RWPM} model, we show how to compute the steady state probabilities for the displacement of users over one-dimensional lattice  after $t\!=\!1$ and $t\!=\!2$ time slots. If the think time is equal to zero, i.e., fast mobility, we also provide an approximation for the displacement probabilities for $t\!>\!2$. We find that the displacement law is location-dependent. 
\item We show how to incorporate the displacement law into the description of the temporal correlation of interference. For the \ac{RWPM} model, we illustrate that the interference correlation is in general higher close to the boundaries, and it decays faster close to the middle of the lattice. With a zero think time, we illustrate how many time slots it takes for the interference to become uncorrelated at different locations.
\item We show how to incorporate the displacement law into the calculation of the conditional outage probability, i.e., the probability of being in outage after $t\!=\!\tau$ time slots, given that an outage occurs at $t\!=\!0$. For the \ac{RWPM} model, we illustrate that a receiver close to the border is unlikely to be in outage, however, the conditional outage probability in the subsequent time slot might be significantly higher than the unconditional. On the other hand, close to the middle of the lattice, the outage events are more probable but also less correlated. 
\end{itemize}

The rest of the paper is organized as follows. In Section~\ref{sec:System} we present the system model, and in Section~\ref{sec:UserDistr} we identify the steady state distribution of users. In Section~\ref{sec:MovementDirection} we compute the displacement probabilities. With these probabilities at hand, we calculate the correlation coefficient of interference in Section~\ref{sec:InterferenceMoments}, and the conditional outage probability in Section~\ref{sec:Outage}. Numerical examples are available in Section~\ref{sec:Numericals}, and conclusions in Section~\ref{sec:Conclusions}. 

\section{System model}
\label{sec:System}
We consider one-dimensional finite lattice of size $N$, $n\!=\!1,2,\ldots N$, and $K$ users moving across the lattice. Initially, the users are allocated uniformly. Then, each user selects uniformly at random a destination point, and travels towards it with a constant speed $v$. The time is discretized in time slots $t\in\mathbb{N}$. For the time being, the user speed is normalized, so that the distance covered in a time slot is equal to the distance between two lattice points. The user updates its location in the beginning of a time slot, and keeps on doing so till it reaches the destination. Then, it pauses and thinks for a number of time slots, $m_i$, selected from the discrete uniform distribution $m_i\in\left\{0,1,2,\ldots, M\right\}\, \forall i$, where $M$ is the maximum think time in time slots. Then, the same procedure is repeated. 

The described mobility model is a modified version of the \ac{RWPM} model in the discrete one-dimensional space and time. In the continuous-domain model~\cite{Paolo2003}, every user may select its speed uniformly at random from an interval $[v_{\textnormal{min}},v_{\textnormal{max}}]$. In Section~\ref{sec:Densification}, we will discuss how one can approximate the user displacement law in the continuous one-dimensional space by densifying the lattice. For a zero think time, $M\!=\!0$, we will show how to densify the lattice at an arbitrary order. In Section~\ref{sec:Numericals}, we will illustrate the impact of a randomized user speed with mean equal to $v$ on the correlation coefficients.

We are interested to quantify the interference correlation at different locations in the steady state. We denote $t\!=\!0$ the time slot when the steady state starts. We place a virtual receiver at $x_p\!=\!n\!+\!c, n\!=\!1,2,\ldots,\left\lceil\frac{N}{2}\right\rceil, c\!\in\!\left[0,1\right)$. We assume a common transmit power level $P_t$ for all the users. The \ac{MAC} scheme is slotted ALOHA where each user, at the beginning of a time slot, decides whether to transmit or not, independently of its own activity at previous time slots, and the activities of others. The transmission probability is $\xi$.  The generated interference at location $x_p$ and time slot $t\!\geq 0$ is 
\begin{equation}
\label{eq:Interference}
\mathcal{I}\left(x_p,t\right) = P_t \sum\nolimits_{i=1}^K {\xi_i(t)\, h_i(t) \, g\left(x_i(t)-x_p\right)},
\end{equation}
where $\xi_i$ is a Bernoulli \ac{RV} describing the $i$-th user activity, $h_i$ is an exponential \ac{RV} modeling Rayleigh fast fading with unit mean $\mathbb{E}\left\{h_i\right\}\!=\!1 \forall i$, $x_i\in \left\{1,2,\ldots, N\right\}$ is a \ac{RV} describing the location of the $i$-th user with \ac{PDF} calculated in the next section, and $g(\cdot)$ is the distance-based propagation pathloss function, $g(x)=\frac{1}{\epsilon+|x|^a}$, where $\epsilon$ is used to avoid singularity at distance $x\!=\!0$ and it normally takes a small positive value. 

In the steady state, the moments of interference become independent of the time $t$ we take the measurements. The Pearson correlation coefficient, $\rho\left(x_p,\tau\right)$,  at time $t=0$, time-lag $\tau$ and point $x_p$ takes the following form  
\begin{equation}
\label{eq:CorrCoeff}
\rho\left(x_p,\tau\right) = \frac{\mathbb{E}\left\{ \mathcal{I}\left(x_p,\tau\right) \mathcal{I}\left(x_p\right)\right\} - \mathbb{E}\left\{ \mathcal{I}\left(x_p\right) \right\}^2} {\mathbb{E}\left\{ \mathcal{I}\left(x_p\right)^2 \right\} - \mathbb{E}\left\{ \mathcal{I}\left(x_p\right) \right\}^2}, 
\end{equation}
where the notation $\mathcal{I}(x_p,0)$ is shortened to $\mathcal{I}(x_p)$. 

While studying the correlation of outage, we will assume that the desired transmitter is placed at a fixed and known distance from the associated  receiver. Unlike the users generating interference, the desired transmitter is active in every time slot at power level $P_t$ with probability one. It is not part of the mobile users generating interference. The location of the desired transmitter is denoted by $x_t$. The channel between the transmitter and the receiver is also subject to unit-mean Rayleigh fading $h_{tx}$, and the distance-based propagation pathloss model is also $g(\cdot)$. The noise power level at the receiver is $P_N$. This kind of transmitter-receiver pair model with fixed mean desired signal strength in a mobile field of interferers has been widely used in the literature, e.g. in~\cite{Gong2014}.

\section{User distribution in the steady state}
\label{sec:UserDistr}
Since the users move independently of each other, it is sufficient to identify the spatial distribution for a single user. Let $\mathcal{N}$ denote the \ac{RV} whose value $n$ is the location of the user at a randomly selected time slot in the steady state. In order to identify the \ac{CDF} of the user location, $\mathbb{P}\left(\mathcal{N}\leq n\right)$, one has to monitor the user for a sufficiently large number of time slots. The CDF is simply equal to the number of time slots spent at the lattice points $\left\{1,2,\ldots,n\right\}$, divided by the number of time slots the user is monitored. 

Let us ignore for the moment the thinking time, i.e., $M=0$, and assume that the user completes $J\to\infty$ travels in the monitoring time. We denote by $T_j$ the number of time slots spent on the $j$-th travel and by $T_{n,j}$ the number of time slots spent at the lattice points $1,2,\ldots,n$ during the $j$-th travel. Then, the \ac{CDF} is
\[
\mathbb{P}\left(\mathcal{N} \leq n \right) \!=\! \lim_{J\to\infty}\frac{\sum\nolimits_{j=1}^J T_{j,n}}{\sum\nolimits_{j=1}^J T_j} =  \frac{\mathbb{E}\left\{ T_n \right\}}{\mathbb{E}\left\{ T \right\}},
\]
where $T$ is a \ac{RV} describing the number of time slots spent on a travel, and $T_n$ is a \ac{RV} describing the number of time slots spent at the lattice points $\left\{1,2,\ldots,n\right\}$ during a travel. 

Since the user speed is fixed and constant, the travel time is proportional to the distance covered. Hence, 
\[
\mathbb{P}\left(\mathcal{N} \leq n \right) \!=\!  \frac{\mathbb{E}\left\{ L_n \right\}}{\mathbb{E}\left\{ L \right\}},
\]
where $L$ is a RV describing the distance, or equivalently, the number of lattice points visited during a travel, and $L_n$ is a RV whose value is equal to the number of lattice points from the set $\left\{1,2,\ldots,n\right\}$ visited during a travel.

In order to compute the expected value of the \ac{RV} $L$, one has to consider all possible travel paths over the lattice and compute their expected length. Assuming that the source and the destination points are different, there are $N(N\!-\!1)$ paths in total. Note that a user selects its destination independent of the source, thus all paths become equally probable. Therefore the expected length of a path is simply the arithmetic average of all the path lengths. Let us denote the source of a path by $s$ and its destination by $d$. The average path length is
\begin{equation}
\label{eq:MeanLeg}
\mathbb{E}\left\{L\right\} \!=\! \frac{2}{N\left(N\!-\!1\right)} \sum\limits_{s=1}^{N-1} 
\sum\limits_{d=s+1}^{N}\!\!{\!\left(d\!-\!s\right)} \!=\! \frac{N\!+\!1}{3}. 
\end{equation} 

Similarly, in order to compute the expected value of the \ac{RV} $L_n$, one may consider all $N(N-1)$ paths, count the number of lattice points from the set $\left\{1,2,\ldots,n\right\}$ that the user visits in each path, and take the arithmetic average.  The computation of $\mathbb{E}\left\{ L_n \right\}$ can be split into four independent cases depending on the relative location of $n,s$ and $d$. Let denote by $L_n^{(j)}, j\!=\!1,2,3,4$ the \ac{RV} describing the number of lattice points from the set $\left\{1,2,\ldots, n\right\}$ visited during a travel for the $j$-th case. 
\begin{itemize}[leftmargin=.25in]
\item $d\!>\!s$ and $d\!\leq\! n$. In that case, the user travels over $\left(d-s\right)$ lattice points which all contribute to the value of $L_n$.
\[
\mathbb{E}\left\{ L_{n}^{(1)}\right\} = \frac{1}{N(N\!-\!1)} \sum\limits_{s=1}^{n-1}\sum\limits_{d=s+1}^n  \!\!{\left(d\!-\!s\right)}.
\]
\item $d\!>\!s$ and $d \!>\! n$. The user initially travels over $\left(n\!-\!s\right)$ points which contribute to the value of  $L_n$ but then, the travel from $(n\!+\!1)$ to $d$ does not make any contribution.  
\[
\mathbb{E}\left\{ L_{n}^{(2)}\right\} \!=\! \frac{1}{N(N\!-\!1)} \sum\limits_{s=1}^n\sum\limits_{d=n+1}^N  \!\!{\left(n\!-\!s\right)}.
\]
\item $d\!<\!s$ and $s \!\leq\! n$. This case is similar to the first case with reversed source and destination points. 
\[
\mathbb{E}\left\{ L_{n}^{(3)}\right\} = \frac{1}{N(N\!-\!1)} \sum\limits_{d=1}^{n-1}\sum\limits_{s=d+1}^n  \!\!{\left(s\!-\!d\right)}.
\]
\item $d\!<\!s$ and $s \!>\! n$. This case is similar to the second case with reversed source and destination points. However, one has to note that both lattice points $n$ and $d$  contribute to the value of the \ac{RV} $L_n$ because the user location is updated in the beginning of a time slot. 
\[
\mathbb{E}\left\{ L_{n}^{(4)}\right\} = \frac{1}{N(N\!-\!1)} \sum\limits_{d=1}^n\sum\limits_{s=n+1}^N \!\!{\left(n\!-\!d\!+\!1\right)}.
\]
\end{itemize}

After computing the expected number of lattice points from the set $\left\{1,2,\ldots, n\right\}$ visited during a travel as the sum of the four individual cases, $\mathbb{E}\left\{ L_n\right\}=\sum\nolimits_{i=1}^4 \mathbb{E}\left\{ L_{n}^{(i)}\right\}$, and dividing it by the expected length of a path $\mathbb{E}\left\{ L\right\}$ computed in equation~\eqref{eq:MeanLeg}, the \ac{CDF} of the user location for the mobile component of the network, $F_m(n)$, can be read as 
\begin{equation}
\label{eq:MoveCompo}
F_m(n) \!=\! \frac{\mathbb{E}\left\{L_n\right\}}{\mathbb{E}\left\{L\right\}}  =  \frac{\left( 3Nn \!-\! 2n^2 \!-\! 1 \right) n}{N(N^2-1)},\,\,\forall n.
\end{equation}

For comparison purposes, the \ac{CDF} of the user location in the continuous-domain \ac{RWPM} model is~\cite{Paolo2003}
\[
F_m(x) =\frac{3x^2}{x_0^2} \!-\! \frac{2x^3}{x_0^3},\,\, 0\leq x \leq x_0. 
\]

We can see that the continuous-domain \ac{CDF} can be obtained as the limit of equation~\eqref{eq:MoveCompo} for a large $N$.

If we consider a positive maximum think time, $M\!>\!0$, the probability $p$ that the user is static at a randomly selected time slot in the steady state is~\cite{Paolo2003}
\[
p = \frac{\mathbb{E}\left\{m\right\}}{\mathbb{E}\left\{m\right\} \!+\! \mathbb{E}\left\{T\right\}} \!=\! \frac{M\!/2}{M\!/2 \!+\! (N\!+\!1)/3}, 
\]
where the index $i$ in the \ac{RV} $m$ has been dropped since the users are indistinct, the term $\frac{M}{2}$ is the expected think time, and the term $\frac{N\!+\!1}{3}$ describes the expected time of a travel. 

Since all the paths are equally probable, the distribution of destination points is uniform. Therefore the distribution of static users is uniform too. The \ac{CDF} of user location can be expressed as the weighted sum of a uniform distribution describing the static component of the network, $F_s(n)\!=\!\frac{n}{N} \forall n$, and the distribution given in~\eqref{eq:MoveCompo} describing the mobile component  
\[
F(n) = p F_s(n) + \left( 1-p \right) F_m(n), \,\, \forall n.
\]

The \ac{PDF} of the user location can be computed as $f(n)\!=\! F(n)\!-\!F(n\!-\!1), n\geq 2$ and $f(1)\!=\!F(1)$. Finally, 
\begin{equation}
\label{eq:SteadyPDF}
f(n) \!=\! \frac{p}{N} \!+\! \left(1\!-\!p\right) \frac{3N\left(2n\!-\!1\right)\!-\!6n\left(n\!-\!1\right)\!-\!3}{N(N^2-1)}, \forall n.
\end{equation}

We note that for a zero think time, in the continuous-domain model, the probability that a user is located exactly at the borders is zero, while in the discrete-domain model, the same probability is $f(1)\!=\!f(N)\!=\!\frac{3}{N(N+1)}$. 

\section{Displacement probabilities}
\label{sec:MovementDirection}
Let us assume that the network has reached the steady state, and the user we monitor is located at the lattice point $n$. Next, we will show how to compute the displacement probability, $\mathbb{P}\left(n+k,\tau\right)$, i.e., the probability that the user moves to the lattice point $(n\!+\!k)$ after $\tau$ time slots. Since the user speed is fixed to one point per time slot, the possible displacement is $k\!\in\!\left\{-\tau,\ldots,0,\ldots,\tau\right\}$. We will show how to compute $\mathbb{P}\left(n\!+\!k,\tau\right)$ for $\tau\!=\!1$ and $\tau\!=\!2$. Also, we will show how to approximate $\mathbb{P}\left (n\!+\!k,\tau\right)$ for $\tau\!>\!2$, assuming a zero think time. 

\subsection{Change of location after $\tau\!=\!1$ time slots}
\label{sec:MovementDirection1} 
The probability that the user does not change its location at  $t\!=\!1$, $\mathbb{P}\left(n,1\right)$, is location-dependent. It can be expressed as the fraction of the static component in the \ac{PDF} given in~\eqref{eq:SteadyPDF} 
\begin{equation}
\label{eq:StaticProb}
\mathbb{P}\left(n,1\right) = \frac{p}{N f\left(n\right)}, \forall n.
\end{equation}

Given that the user changes its location with probability $\left(1\!-\!\mathbb{P}\left(n,1\right)\right)$, the probability it moves to the right can be computed as the fraction of paths crossing the lattice point $n$ while moving to the right, divided by the total number of paths crossing that point. The paths with source  $s\!\in\!\left\{1,2,\ldots, n\right\}$ and destination $d\!\in\!\left\{n\!+\!1,n\!+\!2,\ldots, N\right\}$ cross the point $n$ to the right while the paths with source $s\in\left\{n,n\!+\!1,\ldots, N\right\}$ and destination $d\!\in\!\left\{1,2,\ldots, n\!-\!1\right\}$ cross the point $n$ to the opposite direction. Hence, 
\begin{equation}
\label{eq:MoveProb}
\mathbb{P}\left(n\!+\!1,1\right) \!=\! \frac{\left(1 \!-\! \mathbb{P}\left(n,1\right)\right)  n\left(N\!-\!n\right)}{n(N\!-\!n) \!+\! (n\!-\!1)(N\!-\!n\!+\!1)},  n\!<\!N. 
\end{equation}

Obviously, $\mathbb{P}\left(n\!+\!1,1\right)=0$ for $n\!=\!N$. Also, the probability that the user moves to the left is the complementary probability, $\mathbb{P}\left(n\!-\!1,1\right) \!=\!  1 \!-\! \mathbb{P}\left(n,1\right) \!-\! \mathbb{P}\left(n\!+\!1,1\right)$ for $n\!>\!1$, and $\mathbb{P}\left(n\!-\!1,1\right)\!=\!0$ for $n\!=\!1$.
\begin{remark}[Thinking at the border]
After substituting $n\!=\!1$ into~\eqref{eq:StaticProb}, the probability that a user which is located at the border stays there and thinks at $t\!=\!1$ is  $\mathbb{P}\left(1,1\right)=\frac{M}{M+2}$, thus it is irrespective of the lattice size $N$. 
\end{remark}
\begin{remark}[Thinking at the center]
Let us consider a large lattice. The probability that a user located at the center, $n\!=\!\left\lceil{N\!/2}\right\rceil$, stays there and thinks at $t\!=\!1$ converges to $\lim_{N\to\infty} \mathbb{P} \left(\left\lceil{\frac{N}{2}}\right\rceil,1\right)=\frac{M}{M\!+\!N}$. Thus, for a finite $M$, the user will move with a high probability. Also, starting from~\eqref{eq:MoveProb}, one can show that for a large $N$, the probabilities to move left or right have equal limits, $\lim\limits_{N\to\infty}\mathbb{P}\left(\left\lceil{\frac{N}{2}}\right\rceil\!-\!1,1\right) \!=\! \lim\limits_{N\to\infty} \mathbb{P}\left(\left\lceil{\frac{N}{2}}\right\rceil\!+\!1,1\right) \!=\! \frac{1}{2}\frac{N}{N\!+\!M}$.
\end{remark}

\subsection{Change of location after $\tau\!=\!2$ times lots} 
\label{sec:MovementDirection2} 
Due to the correlated mobility, the displacements at subsequent time slots are not independent. The probability that the user moves to the point $\left(n\!+\!2\right)$  at $t\!=\!2$, $\mathbb{P}(n\!+\!2,2)$,  can be expressed as the sum of the probabilities of two disjoint events: (i) The user travels over a path with source $s\in\left\{1,2,\ldots,n\right\}$ and destination $d\in\left\{n\!+\!2,\ldots,N\right\}$. (ii) The user travels over a path with source $s\in\left\{1,2,\ldots,n\right\}$ and destination  $d\!=\!n\!+\!1$. After reaching the destination at $t\!=\!1$, the user selects a zero think time with probability $\frac{1}{M+1}$ and then, it selects a new destination $d\in\left\{n\!+\!2,\ldots, N\right\}$. Hence, 
\begin{equation}
\label{eq:n2}
\mathbb{P}(n\!+\!2,2) \!=\! \left(1 \!-\! \mathbb{P}(n,1)\right)  \frac{ n\left(N\!-\!n\!-\!1\right)  \!+\! n\frac{1}{M+1}\frac{N\!-\!n\!-\!1}{N\!-\!1}}{n(N\!-\!n) \!+\! (n\!-\!1)(N\!-\!n\!+\!1)}, 
\end{equation}
for $n \!<\! N\!-\!1$, and $\mathbb{P}(n\!+\!2,2) = 0,\,\, n\geq N\!-\!1$. 

In a similar manner, one can compute the probability that the user moves to the point $\left(n\!-\!2\right)$ after two time slots. 
\begin{equation}
\begin{array}{ccl}
\mathbb{P}(n\!-\!2,2) &=& \left( 1 \!-\! \mathbb{P}(n,1) \right)  \Big( \frac{ \left(n\!-\!2\right)\left(N\!-\!n\!+\!1 \right)}{n(N\!-\!n) \!+\!   (n\!-\!1)(N\!-\!n\!+\!1)} + \\ & &  \frac{\left(N\!-\!n\!+\!1\right)\frac{1}{M+1}\frac{n\!-\!2}{N\!-\!1}}{n(N\!-\!n) \!+\! (n\!-\!1)(N\!-\!n\!+\!1)} \Big),
\end{array}
\end{equation}
for $n \!>\! 2$, and $\mathbb{P}(n\!-\!2,2) \!=\! 0, \, n\leq 2$.

The probability that the user moves to the point $(n\!+\!1)$ after two time slots is equal to the sum of the probabilities of the following events: (i) The user travels over a path with source $s\in\left\{1,2,\ldots,n\right\}$ and destination $d\!=\!n\!+\!1$. After reaching its destination, the user selects a nonzero think time with probability $\frac{M}{M+1}$, and thinks over there at  $t\!=\!2$. (ii) The user thinks at $t\!=\!1$ with probability $\mathbb{P}(n,1)$. Then, it selects a new destination $d\in\left\{n\!+\!1,\ldots,N\right\}$. Hence, 
\begin{equation}
\begin{array}{ccl}
\mathbb{P}\left(n\!+\!1,2\right) &=& \left(1\!-\!\mathbb{P}\left(n,1\right)\right) \frac{n \frac{M}{M+1}}{n(N\!-\!n) \!+\! (n\!-\!1)(N\!-\!n\!+\!1)} \,+ \\ [0.3cm]
& &  \mathbb{P}(n,1)\left(1 \!-\! \mathbb{P}(n,2|n,1)\right) \frac{N\!-\!n}{N\!-\!1},
\end{array}
\end{equation}
for $n\!<\!N$, and $\mathbb{P}\left(n\!+\!1,2\right) \!=\! 0, \, n\!=\!N$.

In the above equation, the term  $\mathbb{P}(n,2|n,1)$ describes the conditional probability that a user which thinks at $t\!=\!1$ at the lattice point $n$, keeps on thinking over there at $t\!=\!2$. In order to compute this probability, we use the fact that in the steady state, the fraction of users which are thinking at point $n$ is $\mathbb{P}(n,1)$ at any time slot. At $t\!=\!2$, the users which are thinking at point $n$ can be one of the following types: (i) Users which have been thinking over there at $t\!=\!1$ and $t\!=\!2$. (ii) Users which arrived at point $n$ at $t\!=\!1$, and stay there and think at $t\!=\!2$. In the steady state, the users that arrive at a lattice point are equal, on average, to the users that move from that point to other lattice points, i.e., $\left(1-\mathbb{P}(n,1)\right)$ for lattice point $n$. Out of the users which arrived at point $n$ at $t\!=\!1$, the fraction of users which stay there at $t\!=\!2$ is equal to the fraction of users whose destination is the point $n$, times the probability $\frac{M}{M+1}$ to select a nonzero think time. Hence, 
\[
\mathbb{P}\left(n,1\right) \mathbb{P}\left(n,2 | n,1\right) \!+\! \left(1 \!-\! \mathbb{P}\left(n,1\right)\right) \frac{q M}{M+1} \!=\! \mathbb{P}\left(n,1\right), 
\]
where $q\!=\!\frac{N-1}{n(N\!-\!n) \!+\! (n\!-\!1)(N\!-\!n\!+\!1)}$ is the fraction of paths with destination the point $n$. After solving for $\mathbb{P}\left(n,2 | n,1\right)$ we get 
\begin{equation}
\label{eq:Nested}
\mathbb{P}\left(n,2 | n,1\right) \!=\! 1 \!-\!  \frac{q M}{M+1}\, \frac{1-\mathbb{P}(n,1)}{\mathbb{P}(n,1)}.
\end{equation}

The probability that the user moves to the point $\left(n\!-\!1\right)$ at $t\!=\!2$ can also be expressed in terms of  $\mathbb{P}\left(n,2 | n,1\right)$ 
\begin{equation}
\begin{array}{ccl}
\mathbb{P}\left(n\!-\!1,2\right) &=& \left(1\!-\!\mathbb{P}\left(n,1\right)\right) \frac{\left(N\!-\!n\!+\!1\right) \frac{M}{M+1}}{n(N\!-\!n) \!+\! (n\!-\!1)(N\!-\!n\!+\!1)} \, + \\ [0.3cm]  & & \mathbb{P}(n,1)\left(1 \!-\! \mathbb{P}(n,2|n,1)\right) \frac{n-1}{N-1},
\end{array}
\end{equation}
for $n\!>\!1$, and $\mathbb{P}\left(n-1,2\right) = 0, \, n=1$. 

Finally, the probability that the user is located at the lattice point $n$ after two time slots is equal to the sum of the probabilities of the following  events: (i) The user thinks over there at $t=1$ and $t=2$. (ii) The user travels over a path with source  $s\in\left\{1,2,\ldots,n\right\}$ and destination $d\!=\!n\!+\!1$. After reaching its destination at $t\!=\!1$, the user selects a zero think time with probability $\frac{1}{M+1}$ and then, a new destination $d\in\left\{1,2,\ldots,n\right\}$. (iii) The user travels over a path with source $s\in\left\{n,\ldots,N\right\}$ and destination $d\!=\!n\!-\!1$. After reaching its destination, the user selects a zero think time and then, a new destination $d\in\left\{n,\ldots,N\right\}$. 
\begin{equation}
\label{eq:n0}
\begin{array}{ccl}
\mathbb{P}\left(n,2\right) &=& \left(1-\mathbb{P}(n,1)\right) \frac{\frac{1}{M+1}\frac{ n^2 \!+\! \left(N\!-\!n\!+\!1 \right)^2}{N-1}}{n(N\!-\!n) \!+\! (n\!-\!1)(N\!-\!n\!+\!1)} \, + \\ [0.3cm] & & 
\mathbb{P}(n,1) \mathbb{P}\left(n,2 | n,1\right),
\end{array}
\end{equation} 
for $1\!<\!n\!<\!N$, and $\mathbb{P}\left(n,2\right) \!=\! \frac{(\!1\!-\!\mathbb{P}(n,1)\!)(\!N^2\!+\!1\!)}{(N\!-\!1)^2(M\!+\!1) } + \mathbb{P}(\!n,\!1\!)\mathbb{P}\!\left(n,\!2 | n,\!1\right)$, for $n=\{1,N\}$.

\subsection{Change of location after $\tau\!>\!2$ slots}
\label{sec:HighTau}
The analysis of Section~\ref{sec:MovementDirection2} could be extended to more time slots, $\tau\!>\!2$, however, incorporating all possible user moves in the computation of the displacement probabilities $\mathbb{P}(n\!+\!k,\tau)$ will be cumbersome. One way to get around this issue, is to consider only a limited range of moves, the most probable ones, and obtain approximations for the $\mathbb{P}(n\!+\!k,\tau)$. For instance, the probabilities could be estimated assuming at most one directional change in the user mobility. Even under this assumption, the amount of possible user moves remains high for a large value of time-lag $\tau$ and a positive maximum think time $M.$ Also, the computation of the conditional probabilities, see equation~\eqref{eq:Nested} for $\tau\!=\!2$, becomes highly-nested for $\tau\!>\!2$. Limited by this kind of constraints, we show how to approximate the probabilities $\mathbb{P}(n\!+\!k,\tau)$ only for a zero think time, $M\!=\!0$. 
Studying the long-term correlation of user location under fast mobility, i.e., a high $\tau$ and $M\!=\!0$, can be used as one extreme case,  which will be compared to the other extreme involving no mobility at all. Note that a static network with a uniform density of users can be obtained in the limit of $M\!\to\!\infty$.

\subsubsection{Zero think time}
Let consider a positive $k\!\geq\! 0$, and approximate the probability to move right, $\mathbb{P}(n\!+\!k,\tau),\, n\leq N, \,k \leq N\!-\!n$. The probability to move left can be obtained as $\mathbb{P}(n\!-\!k,\tau)=\mathbb{P}(l\!+\!k,\tau), \,l \!=\! N\!-\!n\!+\!1,\, k \!<\! n$. We start with the special case where $k\!=\!\tau$. The user may reach to the point $(n+\tau)$ in one of the following ways: (i) The user is on a path with source $s\!\in\!\left\{1,2,\ldots, n\right\}$ and destination $d\!\in\!\left\{n+\tau,\ldots, N\right\}$. (ii) The user is on a path with source $s\!\in\!\left\{1,2,\ldots, n\right\}$ and destination $d\!\in\!\left\{n\!+\!1,\ldots,n\!+\!\tau\!-\!1\right\}$. After reaching its destination, the user selects a new destination $d\in\left\{n\!+\!\tau,\ldots, N\right\}$ with probability $\frac{1}{N-1}$. Hence, for $k\!=\!\tau$,
\begin{equation}
\mathbb{P}(n\!+\!k,\tau) = \frac{n\left(N\!-\!n\!-\!\tau\!+\!1\right) \!+\! (\tau\!-\!1) n \frac{N\!-\!n\!-\!\tau\!+\!1}{N\!-\!1}}{n(N\!-\!n)\!+\!(n\!-\!1)(N\!-\!n\!+\!1)}.
\label{eq:Hightau1}
\end{equation}

For $k\!<\!\tau$, the most probable paths ending at location $(n+k)$ after $\tau$ time slots are: (i) The user is on a path with source $s\in\left\{1,2,\ldots, n\right\}$ and destination $d\!= \!n\!+\!k\!+\!\frac{\tau-k}{2}$. After reaching its destination, the user selects a new destination  $d\in \left\{1,\ldots, n\!+\!k\right\}$ with probability $\frac{1}{N-1}$. (ii) In a similar manner, the user first moves to the point $\left(n\!-\!\frac{\tau-k}{2}\right)$, then changes its direction and returns to point $\left(n\!+\!k\right)$. Hence, 
\begin{equation}
\label{eq:Hightauk}
\mathbb{P}(n+k,\tau) = \frac{n \frac{n+k}{N-1} \!+\! (N\!-\!n\!+\!1)\frac{N-n-k+1}{N-1}} {n(N\!-\!n)\!+\!(n\!-\!1)(N\!-\!n\!+\!1)}, 
\end{equation}
for $n\!+\!k\!+\!\frac{\tau\!-\!k}{2} \leq N,  n\!-\!\frac{\tau\!-\!k}{2} \geq 1$. 

When $n\!-\!\frac{\tau\!-\!k}{2} < 1$, only the paths reaching first to $d\!= \!n\!+\!k\!+\!\frac{\tau\!-\!k}{2}$, and then returning to $(n\!+\!k)$ should be considered 
\begin{equation}
\label{eq:Hightauk1}
\mathbb{P}(n+k,\tau) = \frac{n \frac{n+k}{N-1}} {n(N\!-\!n)\!+\!(n\!-\!1)(N\!-\!n\!+\!1)}, 
\end{equation}
for $n\!+\!k\!+\!\frac{\tau\!-\!k}{2} \leq N,  n\!-\!\frac{\tau\!-\!k}{2} \!<\! 1$.

Similarly, when $n\!+\!k\!+\!\frac{\tau\!-\!k}{2} \!>\! N$,  only the paths reaching first to $d\!=\! n\!-\!\frac{\tau-k}{2}$, and then returning to $(n\!+\!k)$ should be counted 
\begin{equation}
\label{eq:Hightauk2}
\mathbb{P}(n+k,\tau) = \frac{(N\!-\!n\!+\!1)\frac{N-n-k+1}{N-1}} {n(N\!-\!n)\!+\!(n\!-\!1)(N\!-\!n\!+\!1)}, 
\end{equation}
for $n\!+\!k\!+\!\frac{\tau\!-\!k}{2} \!>\! N,  n\!-\!\frac{\tau\!-\!k}{2} \!\geq\! 1$. 

Obviously, $\mathbb{P}(n\!+\!k,\tau)\!=\!0$ when $n\!+\!k\!+\!\frac{\tau\!-\!k}{2} \!>\! N$,  and $n\!-\!\frac{\tau\!-\!k}{2} \!<\! 1$. Also, due to the fact that the think time is zero, one has to note that for an even $\tau$, $\mathbb{P}(n\!+\!k,\tau)=0$ for $k\!=\!2l+1, l\in\left\{ -\frac{\tau}{2},-\frac{\tau}{2}\!+\!1,\ldots,\frac{\tau}{2}\!-\!1\right\}$, and similarly, for an odd $\tau$, $\mathbb{P}(n\!+\!k,\tau)\!=\!0$ for $k\!=\!2l, l=\left\{ -\frac{\tau-1}{2},-\frac{\tau-3}{2},\ldots, \frac{\tau-1}{2}\right\}$. 

The approximations for the user displacement can be used to estimate the number of time slots required to see uncorrelated interference at the receiver. In the numerical examples, we will see that the interference becomes uncorrelated for values of $\tau$ significantly smaller than the lattice size. When the time-lag $\tau$ is comparable to the lattice size, the user may change its direction more than once with a non-negligible probability, and the approximation accuracy of equations~\eqref{eq:Hightau1}$-$\eqref{eq:Hightauk2} may degrade. However, in this order of time-lags, the interference correlation is low. 

\subsection{Higher user speeds and continuous approximation}
\label{sec:Densification}
When the user speed is an integer larger than one lattice point per time slot, $v\!>\!1$, the steady state distribution of users with \ac{RWPM} can be obtained from equation~\eqref{eq:SteadyPDF} for $p \!=\! \frac{M\!/2}{M\!/2 \!+\! (N\!+\!1)/\left(3v\right)}$, where $\frac{N\!+\!1}{3v}$ is the expected time of a travel. With user speed $v\!>\!1$, the amount of time needed to travel between adjacent lattice points becomes $dt\!=\!v^{-1}$ time slots, and the user can select uniformly at random between $\left(\frac{M}{dt}+1\right)$ possible pause times. Therefore, the user displacement probabilities with $v\!>\!1$ can be  derived from the displacement probabilities for $v\!=\!1$, keeping in mind that the think time $M$ should be divided with the time discretization interval $\frac{M}{dt}$. The displacement probabilities at time-lag $\tau$ with user speed $v\!=\!1$ are essentially equal (after scaling $M$) to the displacement probabilities at time-lag one with user speed $v\!=\!\tau$, or equal to the displacement probabilities at time-lag two with user speed $v\!=\!\frac{\tau}{2}$, and so forth. Therefore, for a positive think time, the user displacement law at $\tau\!=\!1$ with  speed $v\!=\!2$ is available in equations~\eqref{eq:n2}$-$\eqref{eq:n0} after substituing $M$ by $2M$. In a similar manner, one may use equations~\eqref{eq:Hightau1}$-$\eqref{eq:Hightauk2} to approximate the displacement law for zero think time and $v\!>\!1$.  

\begin{figure}[!t]
 \centering
  \includegraphics[width=3.5in]{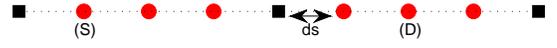}
 \caption{Example illustration of user mobility with $N\!=\!3$ and $N_{\!d}\!=\!4$. Lattice points are the squares and the densification points are the circles. The user speed is $v\!=\!1$ thus, $dt\!=\!\frac{1}{4}$. Source and destination points are indicated by (S) and (D). Travel starts at $t\!=\!0$. The user reaches the destination at $t\!=\!5 dt$. The user location at $t\!=\!4 dt$ is used to compute the correlation at $\tau\!=\!1$.} 
 \label{fig:DenseExample}
\end{figure}

Besides the fixed and common speed for all users, another simplification of the system model in Section~\ref{sec:System} is the discretization of the one-dimensional space and time. A natural way to approximate the user displacement law in the continuous one-dimensional space is to densify the lattice, while keeping the user speed fixed. We denote the densification factor by $N_{\!d}\!\geq\!1$. Obviously, $N_{\!d}$ should be an integer, and $N_{\!d}\!=\!1$ corresponds to no densification. If the distance between adjacent lattice points in the original lattice is normalized to one, in the densified lattice, the distance between adjacent lattice points becomes $ds\!=\!\frac{1}{N_{\!d}}$, and the amount of time needed to cover this distance is $dt\!=\!\frac{ds}{v}$. For an illustration, see Fig.~\ref{fig:DenseExample}. The steady state distribution of users is still  obtained from equation~\eqref{eq:SteadyPDF} after substituting $N$ by $N_{\!d}N$. In addition, $p\!=\!\frac{M\!/2}{M\!/2+\mathbb{E}\left\{T\right\}}$, but the mean travel time should be calculated as $\mathbb{E}\left\{T\right\}\!=\!\frac{N_{\!d}N\!+\!1}{3vN_{\!d}}$. 

For a positive think time, the available displacement probabilities in Section~\ref{sec:MovementDirection2} set a limit on the maximum densification factor, i.e., we can only densify by a factor of $N_{\!d}\!=\!2$ and calculate the user displacement up to $\tau\!=\!1$ time slot for user speed $v\!=\!1$. The user displacement law for $N_{\!d}\!=\!2, v\!=\!1$ and  $\tau\!=\!1$ is available in equations~\eqref{eq:n2}$-$\eqref{eq:n0} after substituing $M$ by $2M$ and $N$ by $2N$. On the other hand, for a zero think time, there is no limitation on the densification order $N_{\!d}$. In Section~\ref{sec:Numericals}, we will approximate the continuous one-dimensional space by densifying the lattice to the order $N_{\!d}\!=\!10^3$ and calculate the correlation coefficients for user speed $v\!=\!1$ and $\tau\!=\!1$. We will also illustrate the impact of randomized user speed, while keeping the mean speed fixed, on the correlation coefficients.  

\section{Interference correlation}
\label{sec:InterferenceMoments}
In order to compute the correlation coefficient of interference, see equation~\eqref{eq:CorrCoeff}, one has to compute the first two moments, and the first-order cross-moment of the interference \ac{RV}. The \ac{MGF} of interference $\Phi_{\mathcal{I}}$ at two time slots $t,\tau$ is 
\[
\Phi_{\mathcal{I}} \!= \!\int\limits\sum\nolimits_{{ {\rm \xi, \rm x}}}{e^{s_{1} \mathcal{I}(t) + s_2\mathcal{I}(\tau)}f_{\!\rm x} \,f_{\xi}\, f_{{\rm h}} \, {\rm d h}}
\]
where $\rm \xi$, $\rm h$ and $\rm x$ are vectors of \acp{RV} with $K$ elements each, $\xi_i, h_i$ and $x_i$ respectively, and $f_{\rm \xi}, f_{\rm h}, f_{\rm x}$ are the joint user \acp{PDF} of activity, fading and location. 

Assuming  \ac{i.i.d.} fading, activity and location among the users, the first moment of interference can be computed from the first derivative of the \ac{MGF} $\frac{\partial \Phi_{\mathcal{I}}}{\partial s_1}$ at $s_1\!=\!0$.
\begin{equation}
\arraycolsep=1.4pt\def\arraystretch{2.2}
\begin{array}{lcl}
\label{eq:FirstMom}
\mathbb{E}\left\{\mathcal{I}\left(x_p\right)\right\} 
&=&P_t \displaystyle \sum\limits_{i=1}^K {\int \sum\limits_{{\rm \xi, x}} \xi_i \,h_i\, g\!\left(x_i-x_p\right)\, f_{\!\rm x} \,f_{\xi}\, f_{{\rm h}}\,{\rm d h}} \\
&=& P_t \displaystyle \sum\limits_{i=1}^K\mathbb{E}\left\{h_i\right\}\, \mathbb{E}\left\{\xi_i\right\}\, \sum\limits_{x_i}{g\!\left(x_i\!-\!x_p\right) f_{x_i}} \\ 
&=& K  \, \xi \, \displaystyle \sum_{n=1}^N {g\!\left(n\!-\!x_p\right) f\!\left(n\right)}.
\end{array}
\end{equation}
where it has been used that $\mathbb{E}\left\{\xi_i\right\}\!=\!\xi$ and  $\mathbb{E}\left\{h_i\right\}\!=\! 1, \, \forall i$, and $P_t\!=\!1$. 
Following the same assumptions, the second moment of interference is 
\begin{equation}
\label{eq:SecondMom}   
\arraycolsep=1.4pt\def\arraystretch{2.2}
\begin{array}{lcl}
\mathbb{E}\left\{\mathcal{I}\left(x_{\!p}\right)^2\right\} 
&=&\displaystyle \sum\limits_{i=1}^K {\int{\sum\limits_{{\rm \xi, x}}{  { \xi_i^2 h_i^2 g\!\left(x_i\!-\!x_p\right)^2 f_{\!\rm x} f_{\xi} f_{{\rm h}} {\rm d h} }}}} + \\
&{}& \!\!\!\!\!\!\!\!\!\!\!\!\!\!\!\!\!\!\!\!\!\!\!\!\!\! \displaystyle \sum\limits_{i=1}^K\sum\limits_{j\neq i} {\int{\sum\limits_{{\rm \xi,x}}{{ \xi_i \xi_j h_i h_j g\!\left(x_i\!-\!x_p\right) g\!\left(x_j\!-\!x_p\right) f_{\!\rm x} f_{\xi} f_{{\rm h}}  {\rm d h}  }}}} \\
&=&\displaystyle \sum\limits_{i=1}^K \mathbb{E}\!\left\{\xi_i^2\right\} \mathbb{E}\!\left\{h_i^2\right\} \sum\limits_{x_i}{g\left(x_i\!-\!x_p\right)^2 \!f_{x_i}} + \\
&{}&  \!\!\!\!\!\!\!\!\!\!\!\!\!\!\!\!\!\!\!\!\!\!\!\!\!\! \displaystyle \sum\limits_{i=1}^K\sum\limits_{j\neq i} \!\mathbb{E}\!\left\{\xi_i \xi_j \right\} \mathbb{E}\!\left\{h_i h_j\right\} \!\sum\limits_{x_i}{\!g\!\left(x_{\!i}\!-\!x_{\!p}\right) \!f_{\!x_{\!i}}} \!\sum\limits_{x_j}{\!g\!\left(x_{\!j}\!-\!x_{\!p}\right) \!f_{\!x_{\!j}}} \\
&=& K \displaystyle \mathbb{E}\left\{\xi^2\right\} \, \mathbb{E}\left\{h^2\right\} \sum\limits_{n=1}^N{g\!\left(n\!-\!x_p\right)^2 \!f(n)} + \\ 
&{}&  \!\!\!\!\!\!\!\!\!\!\!\!\!\!\!\!\!\!\!\!\!\!\!\!\!\! K\left(K\!-\!1\right) \displaystyle \mathbb{E}\left\{\xi\right\}^2 \mathbb{E}\left\{h\right\}^2 \left(\sum_{n=1}^N{g\!\left(n\!-\!x_p\right)}f(n)\right)^2 \\ 
&=& \displaystyle 2 K \xi \!\sum_{n=1}^N{\!g\!\left(n\!-\!x_p\right)^2\!\! f(n)} \!+\!  \frac{K-1}{K}\mathbb{E}\!\left\{\mathcal{I}\!\left(x_p\right)\right\}^{\!2}
\end{array}
\end{equation}
where it has been used that $\mathbb{E}\left\{\xi_i^2\right\}\!=\!\xi$ and $\mathbb{E}\left\{h_i^2\right\}\!=\! 2 \,\forall i$. 

Note that the term $\frac{1}{K}\mathbb{E}\left\{ \mathcal{I}\left(x_p\right)\right\}^2$ in equation~\eqref{eq:SecondMom} essentially describes the difference in the variances of a \ac{PPP} and a \ac{BPP} with the same density of users. The variance of the \ac{BPP} is smaller, because the number of users is fixed. One may argue that the last term becomes negligible for a large $K$, thus propose to approximate the variance of a \ac{BPP} with the variance of the equi-dense \ac{PPP}. While this might be true, one should be aware that the same approximation might induce non-negligible errors in the approximation of the cross-moments of interference. In the numerical examples we will illustrate that for some large $K$, even though the \ac{PPP} provides a good approximation for the variance of interference generated by a \ac{BPP}, it results in significant approximation errors for the correlation coefficients. 

The interference cross-correlation at time-lag $\tau$ can be computed following similar steps to those in equation~\eqref{eq:SecondMom}. One has to note that the vectors  $\rm \xi$, $\rm h$ and $\rm x$ have now $2K$ elements each, describing the activity, fading and location for $K$ users at two different time slots, $t\!=\!0$ and $t\!=\!\tau$, e.g., ${\rm x}\!=\!\left( x_1, x_1^\tau, \ldots, x_K, x_K^\tau\right)^T$, where the notations $x_i(0), x_i(\tau)$ are shortened to $x_i, x_i^\tau$, and $(\,)^T$ denotes the transpose of the vector. Also, let denote by $\rm x_i$ the vector of \acp{RV} describing the locations of the $i$-th user at two time slots, ${\rm x}_i \!=\! \left( x_i,x_i^\tau\right)^T$. The fading and activity are assumed \ac{i.i.d.} over different time slots and users. On the other hand, the locations of a user over time are correlated due to the mobility model. Following the steps detailed on the top of next page we get 
\begin{equation}
\label{eq:CrossCorr}
\mathbb{E}\left\{ \mathcal{I}\!\left(x_p,\tau\right)\! \mathcal{I}\!\left(x_p\right)\right\} \!=\! \displaystyle K \xi^2 \sigma_g(\tau)  \!+\!  \frac{K-1}{K}\mathbb{E}\left\{\mathcal{I}\left(x_p\right)\right\}^2, 
\end{equation}
where the cross-correlation of the distance-based propagation pathoss at time-lag $\tau$, $\sigma_g(\tau)\!\triangleq\! \mathbb{E}_{{\rm x}_i}\!\Big\{g(x_i\!-\!x_p)g(x_i^\tau\!-\!x_p)\Big\}$, is computed after averaging out the user location and displacement
\begin{figure*}[!t]
\normalsize
\[
\arraycolsep=1.4pt\def\arraystretch{2.2}
\begin{array}{lll}
\mathbb{E}\left\{ \mathcal{I}\!\left(x_{\!p},\!\tau\right)\! \mathcal{I}\!\left(x_{\!p}\right)\right\} & = & \displaystyle \sum\limits_{i=1}^K {\!\int{\!\sum\limits_{{\rm \xi,x}}{{ \xi_i\xi_i^\tau h_ih_i^\tau g\!\left(x_i\!-\!x_p\right) g\!\left(x_i^\tau\!-\!x_p\right)\! f_{\!\rm x} f_{\xi} f_{{\rm h}} {\rm d h} }}}} +  \displaystyle \sum\limits_{i=1}^K\sum\limits_{j\neq i} \!{\int{\!\sum\limits_{{\rm \xi,x}}{{ \xi_i \xi_j^\tau h_i h_j^\tau g\!\left(x_i\!-\!x_p\right) g\!\left(x_j^\tau\!-\!x_p\right) \! f_{\!\rm x} f_{\xi} f_{{\rm h}} {\rm d h} }}}} \\
&=& \displaystyle \sum\limits_{i=1}^K \!\mathbb{E}\!\left\{\xi_i\!\right\}^2 \mathbb{E}\!\left\{h_i\!\right\}^2 \!\sum\limits_{{\rm x}_i}\! {g\!\left(x_i\!-\!x_{\!p}\right) \!g\!\left(x_i^\tau\!-\!x_{\!p}\right) \!f_{{\!\rm x}_i}} \!+\! \displaystyle \sum\limits_{i=1}^K\sum\limits_{j\neq i} \!\mathbb{E}\!\left\{\xi_i\xi_j^\tau \right\} \mathbb{E}\!\left\{h_ih_j^\tau\right\}\! \sum\limits_{x_i}{g\!\left(x_i\!-\!x_p\right)\!f_{x_i}}\! \sum\limits_{x_j^\tau}{g\!\left(x_j^\tau\!-\!x_p\right)\!f_{x_j^\tau}}  \\
&=&  \displaystyle \xi^2 \sum\nolimits_{i=1}^K{\sum\nolimits_{{\rm x}_i} g\left(x_i\!-\!x_p\right) g\left(x_i^\tau\!-\!x_p\right) f_{{\rm x}_i}} + 
 K(K\!-\!1) \displaystyle \xi^2 \left( \sum\nolimits_{n=1}^N {g\left(n-x_p\right) f(n)}\right)^2.  
\end{array}
\]
\hrulefill
\end{figure*}
\begin{equation}
\label{eq:Expectg}
\sigma_g(\tau)\!=\!\sum\limits_{n\!=\!1}^N\sum\limits_{k=\!-\tau}^\tau \!g(n\!-\!x_p)  g(n\!+\!k\!-\!x_p) \mathbb{P}(n\!+\!k,\tau) f(n). 
\end{equation}

Equations~\eqref{eq:FirstMom}$-$\eqref{eq:Expectg} are one of the main results of this paper showing how to incorporate the displacement law under any mobility model over one-dimensional lattice, into the interference correlation model. For the mobility model considered in this paper, the displacement probabilities are location-dependent, i.e., a randomly selected user is more probable to be in the middle of a flight when it is located close to the center. After substituting equations~\eqref{eq:FirstMom}$-$\eqref{eq:Expectg} into equation~\eqref{eq:CorrCoeff}, one can compute the correlation coefficient which is location-dependent too. One may also deduce that the correlation coefficient is independent of the number of users $K$. This is true also for the \ac{PPP} approximation~\cite{Koufos2016}.

In the limit of a large think time, $M\!\to\!\infty$, the static component in the \ac{CDF} of the user location, $F_s(n)$, dominates over the mobile component $F_m(n)$, and the user distribution degenerates to uniform, $f(n)\!=\!\frac{1}{N} \forall n$. After replacing $f(n)$ into equations~\eqref{eq:FirstMom}$-$\eqref{eq:Expectg}, and  $x_i\!=\!x_i(\tau)$, the correlation coefficient for infinite think time, $\rho_{\infty}(x_p)$,  becomes 
\begin{equation}
\label{eq:MInf}
\rho_{\infty}\!\!\left(x_p\right) \displaystyle  \!=\!\frac{\frac{K\xi^2}{N}  \sum\nolimits_{n=1}^N{\!g\!\left(n\!\!-\!\!x_p\right)^2} \!\!-\! \frac{1}{K}\mathbb{E}\left\{\mathcal{I}(x_p)\right\}^{2} } {\frac{2K\xi}{N}\sum\nolimits_{n=1}^N{\!\!g\!\left(n\!\!-\!\!x_p\right)^2} \!\!-\! \frac{1}{K}\mathbb{E}\left\{\mathcal{I}(x_p)\right\}^{2}}.
\end{equation}

Therefore in a static network with a uniform density of users, the correlation coefficient does not depend on the time-lag $\tau$, however, it is still location-dependent. It becomes independent of the location, only if the number of users at each time slot varies according to the Poisson distribution. In that case, the second term in the numerator and denominator of equation~\eqref{eq:MInf} would vanish, and the correlation coefficient becomes equal to $\rho_\infty\!=\!\frac{\xi}{2}$. Note that this is also  the correlation coefficient for a static network modeled by a \ac{PPP} in the infinite two-dimensional plane~\cite{Ganti2009}.

\section{Outage probability}
\label{sec:Outage}
Let  $\mathcal{E}_\tau$ describe the outage event at time slot $t\!=\!\tau$ in the steady state. Firstly, we show how to compute the probability of outage at an arbitrarily selected time slot, referred to as the unconditional outage probability and denoted by $\mathbb{P}(\mathcal{E})$. Secondly, in order to examine the impact of interference correlation on the outage correlation, we show how to compute the probability of outage at time slot $t\!=\!\tau$ given that an outage occurs at $t\!=\!0$. This is referred to as the conditional outage probability and it is denoted by $\mathbb{P}(\mathcal{E}_\tau|\mathcal{E}_0)$. 

\subsection{Unconditional outage probability} 
It is assumed that when the \ac{SINR} falls under a target level  $q$, the receiver is not able to decode the desired transmission and an outage occurs. The probability of outage, $\mathbb{P}\left({\text{SINR}}\leq q\right)$, given the location of the desired transmitter at $x_t$ is 
\[
\begin{array}{ccl}
\mathbb{P}(\mathcal{E}) &=& \mathbb{P}\left(\frac{P_t g\left(x_t-x_p\right) h_{tx}}{P_N + \mathcal{I}(x_p)} \leq q \right) \\ [0.3cm]
&=& \mathbb{P}\left(h_{tx}\leq \frac{q\,\left(P_N+\mathcal{I}(x_p)\right)} {P_t g\left(x_t-x_p\right)}\right).
\end{array}
\]

Taking the average over all possible spatial, fading and activity realizations of the interferers, and keeping in mind that the fading in the access link is also Rayleigh we get
\begin{equation}
\label{eq:UncondOutage}
\arraycolsep=1.4pt\def\arraystretch{2.2}
\begin{array}{ccl}
\mathbb{P}(\mathcal{E})&=&\mathbb{E}_{\mathcal{I}(x_p)}\left\{ \mathbb{P}\left(h_{tx}\leq \frac{q\left(P_N+\mathcal{I}(x_p)\right)}{P_t g\left(x_t-x_p\right)} \Bigg \vert \mathcal{I}(x_p) \right) \right\} \\ [0.3cm]
&=& 1 - e^{-\mathcal{P}_N}\mathbb{E}_{\mathcal{I}(x_p)}\left\{ e^{-s\mathcal{I}(x_p)} \right\},
\end{array}
\end{equation}
where $s=\frac{q}{P_t g\left(x_t-x_p\right)}$ and $\mathcal{P}_N=\frac{qP_N}{P_tg(x_t-x_p)}$. 

Let define the discrete-valued function $G(x_k)=\frac{1}{1+sP_tg(x_k-x_p)}$. The Laplace Transform of the interference, $\mathcal{L}_{\mathcal{I}}\!=\! \mathbb{E}_{\mathcal{I}(x_p)}\!\!\left\{ e^{-s\mathcal{I}(x_p)} \right\} \!=\! \mathbb{E}_{x_k,\xi_k,h_k}\!\!\!\left\{ e^{-s P_t\!\sum\nolimits_{k}\xi_kh_kg(x_k-x_p)} \right\}$  may take the following form 
\begin{equation}
\label{eq:UnCondExpect}
\arraycolsep=1.4pt\def\arraystretch{2.2}
\begin{array}{lcl}
\mathcal{L}_{\mathcal{I}} 
&\stackrel{(a)}{=}& \displaystyle  \mathbb{E}_{x_k,\xi_k}\!\!\!\left\{ \prod\nolimits_{k}{\frac{1}{1+sP_t\xi_k g(x_k-x_p)}}\right\}\\ [0.3cm] 
&\stackrel{(b)}{=}& \displaystyle  \mathbb{E}_{x_k}\!\!\!\left\{\! \prod\limits_{k}{\!1\!-\!\xi \!+\! \xi G(x_k)}\!\right\}  
\!\stackrel{(c)}{=}\!  \displaystyle  \left(\! 1\!-\!\xi \!+\! \xi \! \sum\limits_{n=1}^N{\!G(n)f(n) }\!\!\right)^{\!\!K}\!\!\!\! .
\end{array}
\end{equation}

In the above equations, (a) follows from the \ac{i.i.d.} Rayleigh fading in the interfering links which is independent of the locations and activities of the users, (b) follows from averaging over ALOHA, and (c) from the \ac{i.i.d.} locations of interferers. After substituting equation~\eqref{eq:UnCondExpect} into~\eqref{eq:UncondOutage}, the unconditional outage probability becomes
\begin{equation}
\label{eq:ConnFixedTx}
\mathbb{P}(\mathcal{E}) \!=\!  1 \!-\! e^{-\mathcal{P}_N} \left( 1\!-\!\xi \!+\! \xi  \sum\nolimits_{n=1}^N{ G(n) f(n)}\right)^K.
\end{equation}

\subsection{Conditional outage probability}
The conditional probability of outage can be written in terms of the joint probability of outage at $t\!=\!0$ and $t\!=\!\tau$,  $\mathbb{P}\left( \mathcal{E}_\tau|\mathcal{E}_0\right) \!=\! \frac{\mathbb{P}\left(\mathcal{E}_\tau,\mathcal{E}_0\right)}{\mathbb{P}\left(\mathcal{E}\right)}$. The joint probability of outage $\mathbb{P}\left(\mathcal{E}_\tau,\mathcal{E}_0\right) \!=\! \mathbb{P}\left(h_{tx}\leq H_{tx}, h_{tx}(\tau)\leq H_{tx}(\tau)\right)$ is 
\begin{equation}
\label{eq:JointOutage}
\arraycolsep=1.4pt\def\arraystretch{2.2}
\begin{array}{ccl}
\mathbb{P}\left(\mathcal{E}_\tau,\mathcal{E}_0\right) 
&=& \mathbb{E}_{{{\ssmall \mathcal{I}(x_p), \mathcal{I}(x_p,\!\tau)}}}\!\Bigg\{\! \Big(\!1-e^{-\mathcal{P}_N}e^{-s\mathcal{I}(x_p)} \, - \\ [0.3cm] 
& & e^{-\!\mathcal{P}_{\!N}}\!e^{-\!s\!\mathcal{I}(x_p,\tau)} \!\!+\! e^{-\!2\mathcal{P}_{\!N}}e^{-\!s\!\mathcal{I}(\!x_p\!)} e^{-\!s\!\mathcal{I}(x_p,\tau)}\!\Big)\!\Bigg\} \\  [0.3cm] 
& = & 1 - 2(1-\mathbb{P}\left(\mathcal{E}\right)) + e^{-2\mathcal{P}_N}  \mathcal{L}_{\mathcal{I}}(\tau), 
\end{array}
\end{equation}
where $H_{tx}\!=\!\frac{q \left(P_N+ \mathcal{I}(x_p)\right)}{P_tg(x_t-x_p)}$, $ H_{tx}(\tau) \!=\! \frac{q \left(P_N+\mathcal{I}(x_p,\tau)\right)}{P_tg(x_t-x_p)}$, and the joint Laplace functional of the interference at time slots $t\!=\!0$ and $t\!=\!\tau$,  $\mathcal{L}_{\mathcal{I}}(\tau) \!=\! \mathbb{E}_{{{\ssmall \mathcal{I}(x_p), \mathcal{I}(x_p,\!\tau)}}}\!\left\{e^{-s\mathcal{I}(x_p)} e^{-s\mathcal{I}(x_p,\tau)}\right\}$ is 
\begin{equation}
\arraycolsep=1.4pt\def\arraystretch{2.2}
\begin{array}{ccl}
\mathcal{L}_{\mathcal{I}}(\tau) 
&\stackrel{(a)}{=}& \displaystyle \mathbb{E}_{{{ {\rm{x}}_k }}}\!\!\!\left\{ \! \prod\limits_{k} \!1 \!-\! \xi \!+\! \xi G(x_k) \!\prod\limits_k \!1 \!-\! \xi \!+\! \xi G(x_k(\tau)) \!\right\} \\ [0.3cm]
&=&   \displaystyle
\Big(\! \mathbb{E}_{ {\rm{x}}_k}\!\!\left\{ \left(1\!-\!\xi \!\!+\! \xi G(x_k)\right) \left( 1\!-\!\xi\!\!+\! \xi G(x_k(\!\tau\!)) \right) \right\} \!\Big)^{\!K} \\ [0.3cm] 
&=& \displaystyle \Big(\! (1\!-\!\xi)^2\!\!\!+\!2\xi(1\!-\!\xi)\!\!\sum\limits_{n=1}^N\!\!G\!(n)f\!(n) \!+\! \xi^2\sigma_{\!G}(\tau)\!\Big)^{\!K}
\end{array}
\label{eq:ExpectJointInterf}
\end{equation}
where (a) follows using similar steps to equation~\eqref{eq:UnCondExpect}, and the cross-correlation of the function $G(x_k)$ at time-lag $\tau$, $\sigma_G(\tau)\!\triangleq\!\mathbb{E}_{ {\rm{x}}_k}\!\!\left\{G(x_k)G(x_k(\tau))\right\}$, can be computed after averaging over all possible user locations and displacements 
\begin{equation}
\sigma_G(\tau) \!=\! \displaystyle \sum\limits_{n\!=\!1}^N \sum\limits_{k=\!-\tau}^{\tau} \!\!G(n) G(n\!+\!k)\mathbb{P}(n\!+\!k,\tau)f(n).  
\label{eq:ExpectG}
\end{equation}

Equations~\eqref{eq:JointOutage}$-$\eqref{eq:ExpectG} are one of the main results of the paper showing how to incorporate the displacement law over one-dimensional lattice into the joint probability of outage at time-lag $\tau$. For the mobility model considered in the paper and $\tau\!=\!1$, the joint probability of outage becomes 
\begin{equation}
\label{eq:JointConn1FixedTx}
\arraycolsep=1.4pt\def\arraystretch{2.2}
\begin{array}{ccl}
\mathbb{P}\left(\mathcal{E}_1,\mathcal{E}\right)  & = & 2\mathbb{P} \left(\mathcal{E}\right) \!-\! 1 \!+\! e^{-2\mathcal{P}_N} \Bigg( (1\!-\!\xi)^2 \,+ \\ [0.3cm] 
& &  \displaystyle 2\xi(1\!-\!\xi)\!\! \sum\limits_{n=1}^N\!\!G(n)f(n) \!+\! \xi^2\!\!\sum\limits_{n=1}^N\!\! G(n)f(n) \Big( \\  
{}&{}& \,\,\,\,\,\,\,\, \displaystyle 
G(n)\mathbb{P}(n,1) \!+\! G(n\!+\!1) \mathbb{P}(n\!+\!1,1) \, + \\ [0.3cm] & &  G(n\!-\!1) \mathbb{P}(n\!-\!1,1) \Big) \displaystyle \Bigg)^K\!\!. 
\end{array}
\end{equation}

The conditional probability of outage $\mathbb{P}(\mathcal{E}_1|\mathcal{E})$ is computed after dividing equation~\eqref{eq:JointConn1FixedTx} by equation~\eqref{eq:ConnFixedTx}. 
\begin{figure}[!t]
 \centering
  \includegraphics[width=3.5in]{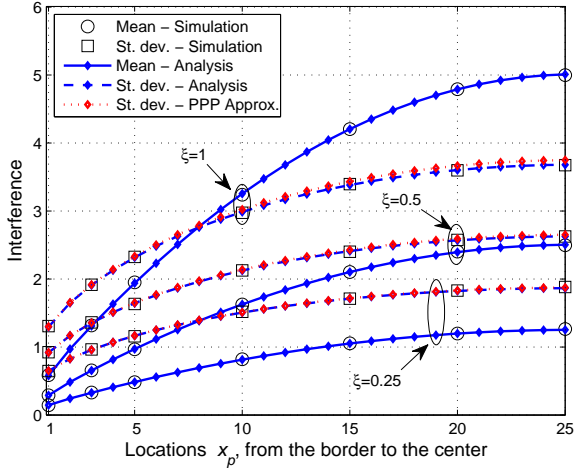}
 \caption{Mean and standard deviation of the interference at different points of the lattice, $x_p\!=\!n,\, n\!=\!1,2,\ldots,\left\lceil\frac{N}{2}\right\rceil$. Pathloss exponent $a\!=\!4$, maximum think time  $M\!=\!5$ time slots and user speed $v\!=\!1$ point per time slot. The rest of the parameter settings can be found in the beggining of Section~\ref{sec:Numericals}.} 
 \label{fig:MeanStd}
\end{figure}

\section{Numerical examples}
\label{sec:Numericals}
We consider a lattice of size $N\!=\!50$ and $K$ users moving according to the \ac{RWPM} model described in Section~\ref{sec:System}. Since the correlation coefficient of interference $\rho$ is independent of the user density, we fix $K\!=\!50$. Initially, we let the network run for $10\,000$ time slots to converge to its  stationary state. Alternative methods to obtain the stationary user distribution are discussed in~\cite{Navidi2004}. The distance-based propagation pathloss model is parameterized with  $\epsilon\!=\!0.5$. 

The validation of the calculation of the mean and standard deviation of interference, see equations~\eqref{eq:FirstMom} and~\eqref{eq:SecondMom}, is illustrated in Fig.~\ref{fig:MeanStd} for activity probabilities $\xi\!=\!1, \xi\!=\!\frac{1}{2}$, and $\xi\!=\!\frac{1}{4}$. While the mean interference is proportional to $\xi$, note that the variance is not, unless the \ac{PPP} approximation is used. As expected, receivers close to the center of the lattice are exposed to higher interference. Also, we see that approximating the \ac{BPP} by the equi-dense \ac{PPP} slightly overestimates the standard deviation. Since the number of users is high, neglecting the  term $\frac{1}{K}\mathbb{E}\left\{ \mathcal{I}\left(x_p\right)\right\}^2$ in equation~\eqref{eq:SecondMom} results in a small approximation error. The error  $\frac{1}{K}\mathbb{E}\left\{\mathcal{I}(x_p)\right\}^2$  is proportional to $\xi^2$, thus the \ac{PPP} approximation may be used in wireless networks with low-traffic, e.g., in sensor networks. 
\begin{figure}[!t]
 \centering
  \includegraphics[width=3.5in]{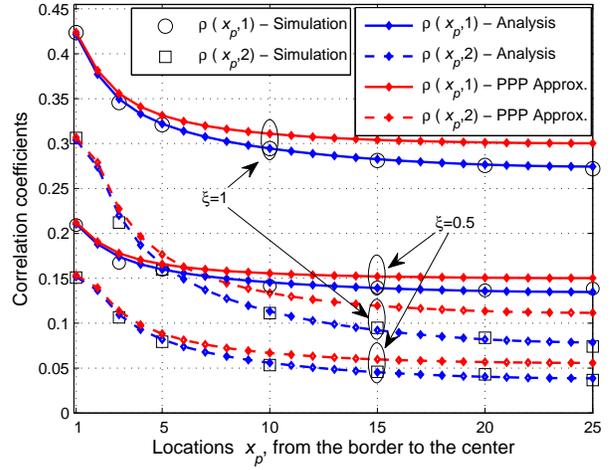}
 \caption{Correlation coefficients for $\tau\!=\!1$ and $\tau\!=\!2$ at different points of the lattice. Same parameter settings used to generate Fig.~\ref{fig:MeanStd}.}
 \label{fig:TempCorr}
\end{figure}

The computation of the correlation coefficient for $\tau\!=\!1$ and $\tau\!=\!2$ is validated in Fig.~\ref{fig:TempCorr}. For the parameter settings used to generate Fig.~\ref{fig:TempCorr}, the correlation is higher at the border than at the center. Also, even within a single time slot, the interference correlation reduces rapidly, particularly close to the center. In order to explain this behaviour, we  recall that close to the boundary a higher fraction of users thinks, see equation~\eqref{eq:StaticProb}, making the interference pattern more correlated over there. Also, users may approach or leave the center from either direction while, close to the boundary, one of the probabilities $\mathbb{P}(n\!+\!1,1),\mathbb{P}(n\!-\!1,1)$ will dominate the other. This means that the user distribution changes more rapidly close to the center, making the interference pattern uncorrelated within few time slots. 

In Fig.~\ref{fig:TempCorr}, we see that the \ac{PPP} approximation  results in non-negligible errors for the correlation coefficients. The \ac{PPP} overestimates both the variance and the first-order cross-moment, see equations~\eqref{eq:SecondMom} and~\eqref{eq:CrossCorr}. The absolute error is the same, $\frac{1}{K}\mathbb{E}\left\{\mathcal{I}(x_p)\right\}^2$, but it affects more the cross-moment, resulting in an overestimation of the correlation coefficients. The \ac{PPP} approximation might not be reliable for computing the temporal correlation of interference generated by a \ac{BPP}. 

\begin{figure}[!t]
 \centering
  \includegraphics[width=3.5in]{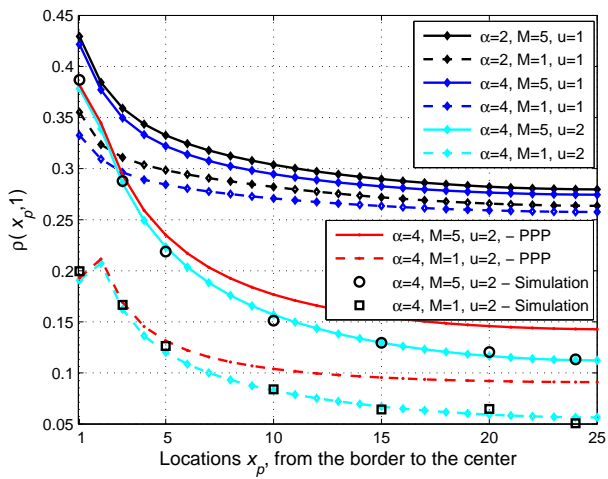}
 \caption{Correlation coefficient $\rho(x_p,1)$ for different think time $M$, user speed $v$ and propagation pathloss exponent $a$. User activity $\xi\!=\!1$. The rest of the parameter settings are the same used to generate Fig.~\ref{fig:MeanStd}. The calculations for user speed $u>1$ are also validated.}
 \label{fig:TempCorrPathloss}
\end{figure}

Next, we study the impact of propagation pathloss exponent, maximum think time and user speed on the correlation coefficient for $\tau\!=\!1$. A lower pathloss exponent reduces the impact of dominant interferers on the interference pattern. Therefore the local interference feature, i.e., mobility of dominant interferers, does not affect the interference profile as much as in the case of a high pathloss exponent. Because of that, lower pathloss exponents are associated with higher interference correlation, see Fig.~\ref{fig:TempCorrPathloss}. In addition, a higher think time and/or a lower user speed makes the network more static, thus increasing the interference correlation too. The \ac{PPP} approximation error is non-negligible also for higher user speeds.

\begin{figure}[!t]
 \centering
  \includegraphics[width=3.5in]{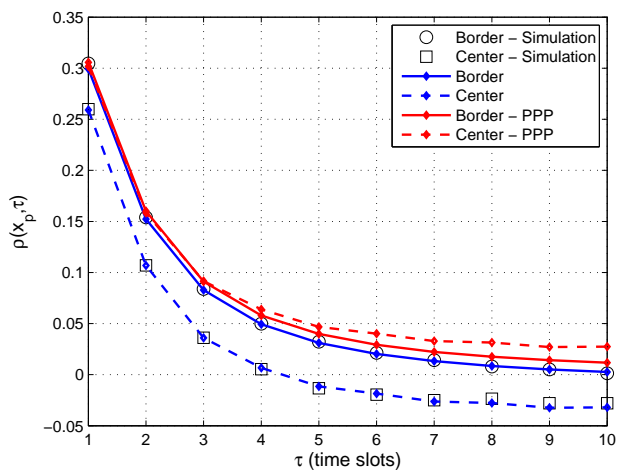}
 \caption{Correlation coefficient $\rho(x_p,\tau)$ with respect to the time-lag $\tau$ at the border of the lattice, $x_p\!=\!1$, and at the center, $x_p\!=\!N\!\!/2$, for zero think time, $M\!=\!0$, pathloss exponent $a\!=\!2$, and user activity $\xi\!=\!1$. The rest of the parameter settings are the same used to generate Fig.~\ref{fig:MeanStd}.}
 \label{fig:HighTau}
\end{figure}

In Fig.~\ref{fig:HighTau}, we examine how long it takes for the interference to become uncorrelated at the border and at the center for zero think time, see Section~\ref{sec:HighTau}. Due to the fact that the user distribution changes rapidly at the center, it takes only four time slots to see uncorrelated interference over there, while at the boundary, $10$ time slots are needed. When the pathloss exponent is higher, $a\!=\!4$, the correlation becomes zero after three time slots at the center and four time slots at the boundary. For a larger lattice, $N\!=\!100$, it takes six time slots at the center and $16$ time slots at the boundary.  Even though the user mobility is correlated, the interference becomes uncorrelated in few time slots; the users have to travel only within a part of the lattice before the correlation becomes negligible. For the parameter settings used to generate Fig.~\ref{fig:HighTau}, we see that the \ac{PPP} approximation does not capture the fact that the correlation at the center should be smaller than at the border. 

In Fig.~\ref{fig:HighTau} we see that small negative correlation coefficients may arise. For time-lags $\tau\!>\!10$, the correlation coefficients will oscillate between small negative and positive values, and progressively converge to zero for $\tau\!>\!100$. This behaviour cannot be captured by the model in Section~\ref{sec:HighTau}. Nevertheless, the model gives an accurate estimate about the amount of time needed and subsequently about the distance covered before the interference correlation becomes negligible. From Fig.~\ref{fig:HighTau}, we may deduce that the approximations for the user displacement probabilities in equations~\eqref{eq:Hightau1}$-$\eqref{eq:Hightauk2} introduce only small errors in the calculation of interference correlation. Actually, the approximation slightly underestimates the actual probabilities for the user displacement because it does not account for all the paths reaching at the point $(n\!+\!k)$ after $\tau$ time slots. As a result, the cross-correlation in equation~\eqref{eq:CrossCorr}, and subsequently the correlation coefficient are slightly underestimated too.  
\begin{figure}[!t]
 \centering
  \includegraphics[width=3.5in]{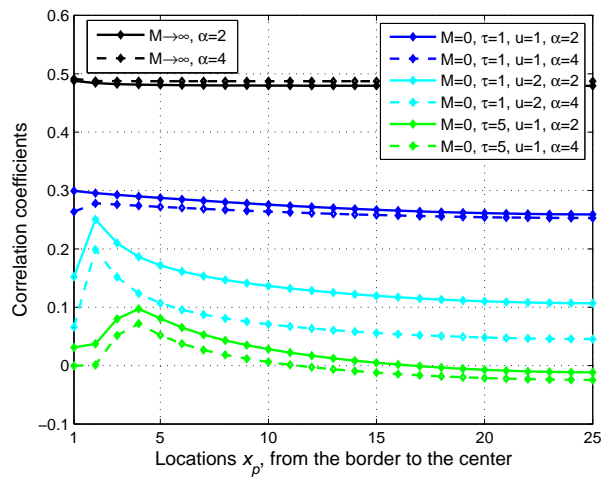}
 \caption{Correlation coefficients in static networks with uniform density of users, $M\!\to\!\infty$, and with \ac{RWPM} for zero think time. User activity $\xi\!=\!1$. The rest of the parameter settings are the same used to generate Fig.~\ref{fig:MeanStd}.}
 \label{fig:MinfM0}
\end{figure}
\begin{figure}[!t]
 \centering
  \includegraphics[width=3.5in]{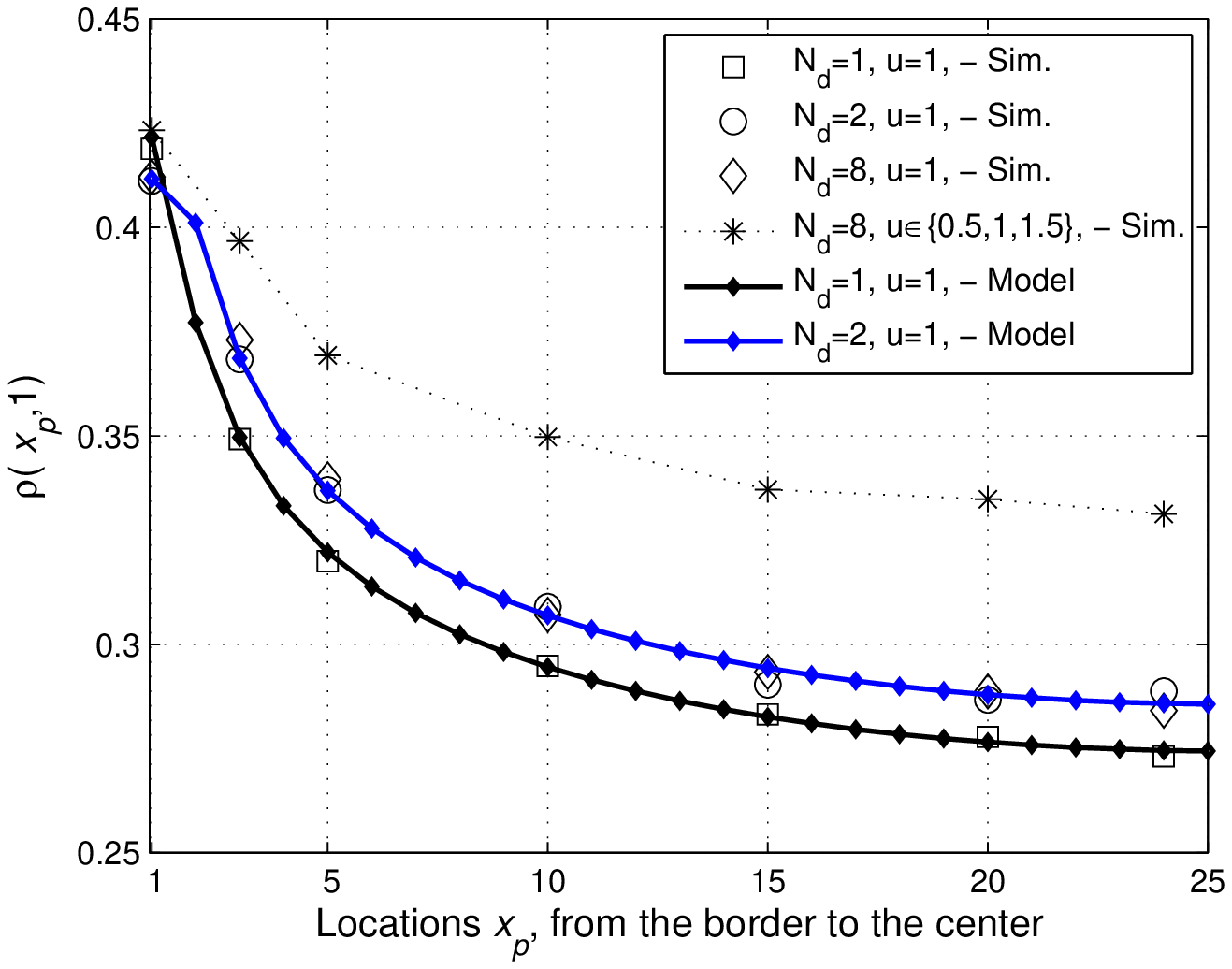}
 \caption{Correlation coefficients at time-lag $\tau\!=\!1$ with different densification factors $N_d$ and positive think time. The parameter settings are the same used to generate Fig.~\ref{fig:MeanStd}.}
 \label{fig:DensificationM5}
\end{figure}
\begin{figure}[!t]
 \centering
  \includegraphics[width=3.5in]{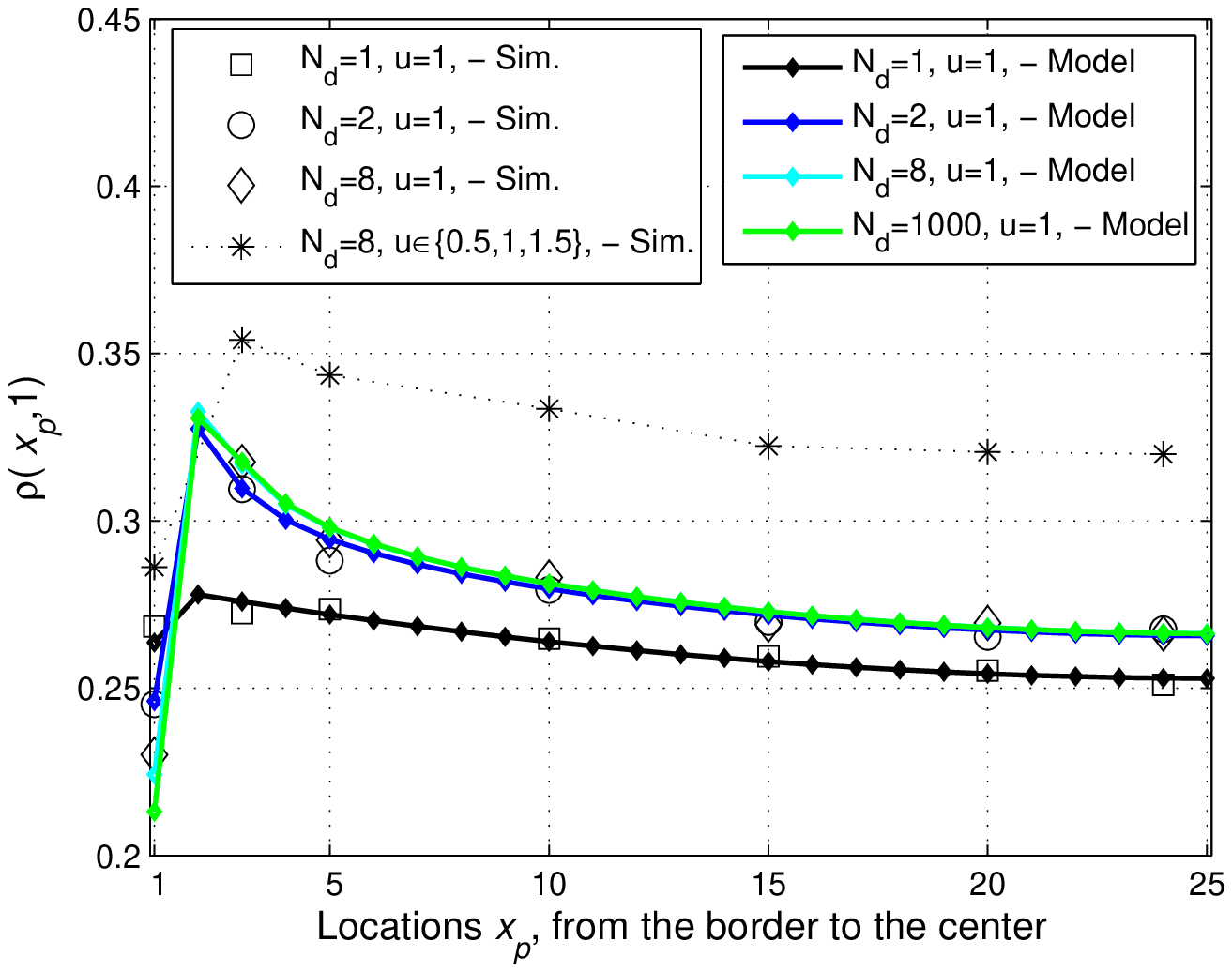}
 \caption{Correlation coefficients at time-lag $\tau\!=\!1$ with different densification factors $N_d$ and zero think time. The parameter settings are the same used to generate Fig.~\ref{fig:MeanStd}.}
 \label{fig:DensificationM0}
\end{figure}
\begin{figure}[!t]
 \centering
  \includegraphics[width=3.5in]{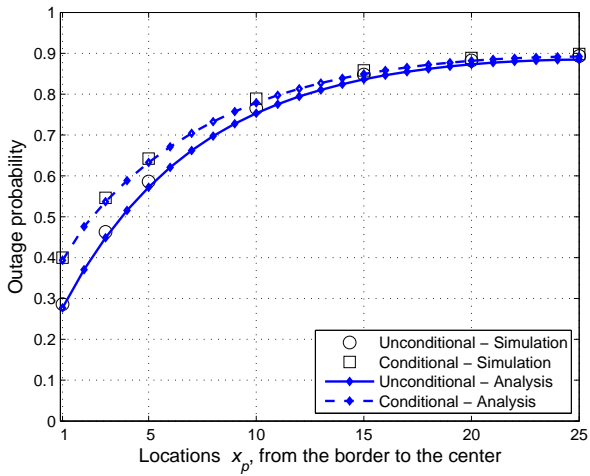}
 \caption{Conditional, $\mathbb{P}(\mathcal{E})$, and unconditional outage probability, $\mathbb{P}(\mathcal{E}_1|\mathcal{E})$, assuming the desired transmitter is located at $x_t\!=\!x_p$. \ac{SINR} target $q\!=\!1$, noise power level $P_N\!=\!10^{-3}$, and propagation pathloss exponent $a\!=\!2$. The rest of the parameter settings are the same used to generate Fig.~\ref{fig:MeanStd}.}
 \label{fig:UncondOutProb}
\end{figure}

In Fig.~\ref{fig:MinfM0}, we study interference correlation under two extreme scenarios, \ac{RWPM} with zero think time, $M\!=\!0$, and static users with a uniform density, $M\!\to\!\infty$. For $M\!\to\!\infty$, the correlation coefficient is calculated in equation~\eqref{eq:MInf}, and it is slightly higher at the boundaries as compared to the center, see Fig.~\ref{fig:MinfM0}. On the other hand, with \ac{RWPM}, the correlation is clearly location-dependent. In addition, the propagation pathloss exponent does not seem to affect much the correlation in either scenario. When there is no correlation in the user traffic and the fading channel, mobility is the key factor reducing the interference correlation, see Fig.~\ref{fig:TempCorrPathloss}$-$Fig.~\ref{fig:MinfM0}. 

For a positive think time $M\!>\!0$, the model in Section~\ref{sec:Densification} allows to densify the lattice only by a factor of $N_d\!=\!2$ for user speed $v\!=\!1$. This model is validated in Fig.~\ref{fig:DensificationM5} for think time $M\!=\!5$. As compared to the case without densification, $N_d\!=\!1$, the higher correlation coefficients for $N_d\!=\!2$ are attributed to the fact that some interferers are located  in-between the lattice points, see Fig.~\ref{fig:DenseExample}. These interferers dominate the interference level. When a dominant interferer moves within time-lag $\tau\!=\!1$ from the location $\left(n-\frac{1}{2}\right)$ to $\left(n+\frac{1}{2}\right)$ the mean interference it generates will remain the same. Therefore, apart from the boundary, the cross-correlation of interference will increase in the densified lattice. In Fig.~\ref{fig:DensificationM5}, we have also carried out simulations with densification factor $N_d\!=\!8$ to resemble more closely the continuous one-dimensional space. In the simulations with $N_d\!=\!8$, we have also considered randomized user speeds, i.e.,  before starting a travel, every user selects its speed uniformly at random from the set $v\in\left\{0.5,1,1.5\right\}$ thus, the mean user speed is still equal to one lattice point per time slot. We see that the model for $N_d\!=\!2$ provides an accurate approximation of the correlation coefficients also for higher densification orders. We have also validated the model in Section~\ref{sec:Densification} for think time $M\!=\!1$ and pathloss exponent $a\!=\!2$; similar approximation accuracy was observed. In addition, when the user speed is randomized, the users are progressively trapped to lower user speeds. This is a well-known feature of the \ac{RWPM} model, see for instance~\cite{Pratt2015}, and it is confirmed in our simulations in Fig.~\ref{fig:DensificationM5}, where we see higher correlation coefficients for randomized user speed.

The model in Section~\ref{sec:Densification} allows for arbitrary densification order only for a zero think time. We consider the densification factor $N_d\!=\!1000$ as a good-enough approximation to the continuous one-dimensional space. The model validation is illustrated in Fig.~\ref{fig:DensificationM0}. Apart from the boundary, the curves for $N_d\!=\!8$ and $N_d\!=\!1000$ practically overlap. Even the densification factor $N_d\!=\!2$ estimates closely the continuous domain approximation. We conclude that the lattice can capture the location-dependent property of interference correlation, however, some lattice densification would be needed to estimate more closely the correlation coefficients in the continuous space.

We have so far seen that the interference profile is in general location-dependent. Under the \ac{RWPM} model, receivers close to the boundaries experience less interference, but the interference correlation is also higher over there. Because of that, when the location of the desired transmitter is fixed and known, the outage probability becomes lower close to the boundary but at the same time, the conditional outage probability is clearly higher than the unconditional, see Fig.~\ref{fig:UncondOutProb}. On the other hand, close to the center, the outage events in subsequent time slots are essentially independent. 

\begin{figure}[!t]
 \centering
  \includegraphics[width=3.5in]{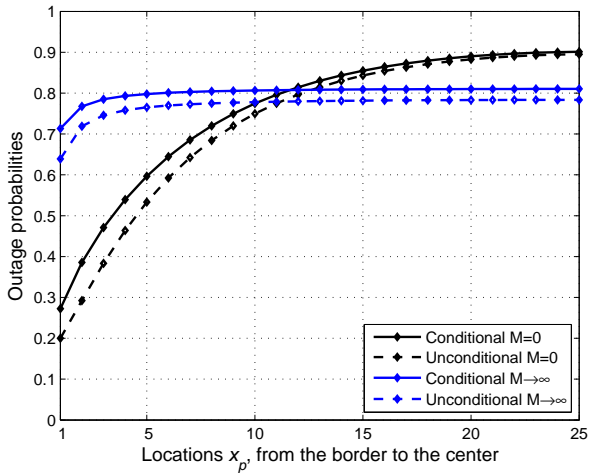}
 \caption{Conditional,  $\mathbb{P}(\mathcal{E})$, and unconditional outage probability, $\mathbb{P}(\mathcal{E}_1|\mathcal{E})$, in highly moving, $M=0$, and static networks, $M\to\infty$. The rest of the parameter settings are the same used to generate Fig.~\ref{fig:UncondOutProb}.}
 \label{fig:CompareStatic}
\end{figure}

Finally, in Fig.~\ref{fig:CompareStatic}, we assess the outage probability under the two extreme cases. In the static case with uniform distribution of users, $M\!\to\!\infty$, the interference and subsequently the outage are still lower close to the border, however, the variations with respect to the  location are smoother. The interference correlation is reduced only due to the randomness in the channel because the users transmit continuously, $\xi\!=\!1$, and they are also static. Therefore, the conditional outage probability becomes clearly higher than the unconditional at all locations.  In the Appendix, we show that for $M\!\to\!\infty$, the conditional outage probability is higher than the unconditional $\mathbb{P}(\mathcal{E}_\tau| \mathcal{E})\!\geq\! \mathbb{P}(\mathcal{E})$ at all locations and time-lags. On the other hand, for $M\!=\!0$, the distribution of users is highly non-uniform, making the interference and outage location-dependent, while the mobility helps reduce the interference correlation particularly close to the center. 

\section{Conclusion}
\label{sec:Conclusions} 
In practice, wireless networks have finite boundaries, and the user mobility is not a random walk.  In order to capture the impact of user mobility over a bounded one-dimensional area on the interference and outage correlation, one has to identify the user distribution and the user displacement law. As an example case, we used the \ac{RWPM} model, and we identified its displacement law over one-dimensional finite lattice. We  illustrated that the temporal correlation of interference is  location-dependent, being higher close to the boundary, where the level of mobility is lower than near the center. Because of that, the interference needs more time to become uncorrelated at the boundary than at the center. Neglecting the particular features of deployment and  user mobility  may lead to erroneous calculation for the interference. In the future, it is important to study exactly how the correlation properties of interference and outage may impact network performance metrics, e.g., local delay, packet travel times over multiple hops, etc. at different locations.


\appendix  
For a uniform user distribution, $f(n)\!=\!\frac{1}{N}\, \forall n$, the outage probability in equation~\eqref{eq:ConnFixedTx} can be written as 
\begin{equation}
\label{eq:App1}
\mathbb{P}(\mathcal{E}) \!=\! 1 \!-\! e^{-\mathcal{P}_N} \left( 1\!-\!\xi \!+\! \frac{\xi}{N}\sum\limits_{n=1}^N{ G(n)}\right)^K.
\end{equation}

Also, for a static network, $\mathbb{P}(n,1)\!=\!1\, \forall n$, the cross-correlation in equation~\eqref{eq:ExpectG} is simplified to 
\begin{equation}
\label{eq:App2}
\mathbb{E}_{ {\rm{x}}_k}\!\!\left\{G(x_k)G(x_k(\tau))\right\} \!=\! \displaystyle \frac{1}{N}\sum\limits_{n\!=\!1}^N G^2(n). 
\end{equation}
After substituting equation~\eqref{eq:App2} into~\eqref{eq:ExpectJointInterf} we get 
\begin{equation}
\label{eq:App3}
\arraycolsep=1.4pt\def\arraystretch{2.2}
\begin{array}{lcl}
\mathcal{L}_{\mathcal{I}}(\tau) &=& 
\displaystyle \Big(\! (1\!-\!\xi)^2\!\!+\!2\frac{\xi(1\!-\!\xi)}{N}\!\!\sum\limits_{n=1}^N\!\!G(n) \!+\!  \frac{\xi^2}{N} \!\!\sum\limits_{n=1}^N \!\!G^2\!(n)\!\Big)^{\!\!K} \\ [0.3cm] 
&\stackrel{(a)}{\geq}& \displaystyle \Bigg( (1\!-\!\xi)^2\!+\!2\frac{\xi(1-\xi)}{N}\sum\limits_{n=1}^N\!G(n) \, + \\ [0.3cm] & & \,\,\,\,\, \displaystyle \frac{\xi^2}{N^2} \left(\sum\limits_{n=1}^N G(n)\right)^2  \Bigg)^K \\ [0.3cm] 
&=& \displaystyle \left( 1\!-\!\xi\!+\!\frac{\xi}{N}\sum\limits_{n=1}^N\!G(n)\right)^{\!\!2K}.
\end{array}
\end{equation}

Here, (a) follows from the Cauchy-Schwarz inequality. Using~\eqref{eq:App3}, the joint outage probability in equation~\eqref{eq:JointOutage} can be upper-bounded as 
\begin{equation}
\label{eq:LemmaJoint2}
\arraycolsep=1.4pt\def\arraystretch{2.2}
\begin{array}{ccl}
\mathbb{P}(\mathcal{E}_\tau,\mathcal{E}) & \geq &  \displaystyle 2\mathbb{P}(\mathcal{E}) \!-\! 1 \!+\! \left(\!\!e^{-\mathcal{P}_N}\!\!\left( \!1\!-\!\xi\!+\!\frac{\xi}{N}\!\!\sum\limits_{n=1}^N\!\!G(n)\!\right)^{\!\!K}\right)^{\!\!\!2} 
\\ [0.3cm] 
&\stackrel{(a)}{=}& 2\mathbb{P}(\mathcal{E}) - 1  + \left( 1 - \mathbb{P}(\mathcal{E})\right)^{2} = \mathbb{P}(\mathcal{E})^2,
\end{array}
\end{equation}
where (a) follows from~\eqref{eq:App1}. As a result, the conditional outage probability is always higher than the unconditional $\mathbb{P}(\mathcal{E}_\tau|\mathcal{E}) \!=\! \frac{\mathbb{P}(\mathcal{E}_\tau,\mathcal{E})}{\mathbb{P}(\mathcal{E})} \!\geq\! \mathbb{P}(\mathcal{E})$. 


%


\ifCLASSOPTIONcompsoc
  \section*{Acknowledgments}
\else
  \section*{Acknowledgment}
\fi
This work was supported by the EPSRC grant number EP/N002458/1 for the project Spatially Embedded Networks.

\ifCLASSOPTIONcaptionsoff
  \newpage
\fi

\end{document}